Multicurrency advisor based on the NSW model. Detailed description and perspectives

**A.M. Avdeenko**

National University of Science and Technology, Moscow, Russia

**1. Introduction**

Previous works have brought about [1-3] great response and some incomprehension of the results obtained. Partially, it is fault of the author's striving to present the material in a possibly compact form without dealing much with details and peculiarities of the algorithms. Also, it is important that financial mathematics is not the science the natural scientists are used to – it has somewhat different method of proof and strictness, and frequently anything, which seems to be natural in the theoretical physics, or in the theory of non-linear differential equations, is not known or is not too popular in the financial statistics.

As well, there are some difficulties with respect to terminology: incorrect, or not fully legal, иor too broad use of some terms having clear enough and strict interpretation. The most unlucky happened to be such an ordinary notion as fractal, etc.

Mention should be made that complications of problems connected with financial forecasts have rather serious grounds. Apart from having very complicated structure featuring multiple nonlinearities and feedbacks, financial system interacts with real economy segment, whose behavior, though connected with the financial component, in many respects is determined by its own peculiarities: cyclicity, limited resources, successes and problems of the scientific and technical progress, political, cultural, and educational components, influences exerted by world market, etc.

With this, according to the author's opinion, solution of the problem is significantly facilitated by the fact, that the problem does not consist in modeling of complicated natural process, but only modeling of the decision making process of the Forex market subjects possessing the same input information, together with comparable computational and analytical possibilities aimed at rather definite and, generally speaking, simple objective.

Great physicist A. Einstein once said: «Everything should be made as simple as possible, but not simpler».

This was the principle, as well as the principle of maximum accessibility for non-mathematicians, which governed the author when writing this work.

**2. Financial series: deterministic chaos or stochasticity? Description space**

Further we will consider financial sequences in discrete time representation. Let us confine ourselves, for purposes of for definiteness, to the Forex market. For the Forex market it is usually assumed that at the $t_n$ time the following values are known : opening prices, closing prices, minimum and maximum price values, prices and volumes of transactions for formation of time frame with period, as well as current (tick-by-tick) value of currency pairs quotations and, thereby, tick-by-tick value for all time interval $t_0, t_1...t_n$ .

For beginning let us confine ourselves to consideration of parameters pertaining to the formed time frame, and let us construct from them any linear combination, e.g. half-sum of the opening – closing prices, or some combinations of the opening and closing prices, and maximum and minimum for this time frame.

This discrete time series $X_n$ in future will be called initial series. It is significant that we can simultaneously construct several such series for all or part of currency pairs traded at the Forex market. That is actually we have a discrete vector series.

Subsequent analysis requires setting of basic hypothesis and exact formulation of the problem being solved. Let us formulate the problem as follows: firstly, at what time it is necessary to enter long or short position (i.e. to buy or to sell currency pair), and secondly, in case initial deposit of a preset value is available, how it should be distributed among various currency pairs to attain maximum trade efficiency, which will be understood as maximum current profit with moderate risk [ 2 ].

Initially, for the purpose of simplicity, we will consider an unidimensional sequence. As a basic hypothesis let us assume that all (here I underline it - all) information necessary for making of efficient decision is contained in the quotations history, that is in the initial sequence $X_n$. By the way, thereby we reject necessity to use so-called fundamental analysis when making the decision.

The last statement is of course debatable, but since we are going to construct exact or maximally quantitative decision making procedure, this assumption has very serious grounds. Firstly, the fundamental analysis results can hardly be evaluated qualitatively, and it is difficult to range weight and evaluate contribution of each of them, secondly, even when they have any grounds, their effect extends to essentially more broad forecasting horizons than that necessary for the decision making, and finally, the fundamental analysis data will in that our other measure influence current quotations or input variables, which, in one way or another, will be analyzed.

In other words, by letting the market himself to «digest» the fundamental forecasts, we will make decisions on the grounds of available information. The same pertains to the so-called insider information: if it is significant, then efficient algorithm of analysis of financial series will detect it and will use it, if it is not – then will not. Moreover, in contemporary world one ceases to feel difference between information and purposeful disinformation, if such disinformation is capable, in that or other sense, to affect making of decision and to turn it to needed direction.

Having input information vector, we can construct description space of the analyzed system. Successful selection of the description space will allow to find out, with minimum losses, the laws governing its behavior. As the description space element we may select certain function based on elements $X_n$. For instance, we can take $Z_n = \ln \dfrac{X_n}{X_{n-1}}$. It is interesting, that statistics $Z_n$ for all currency pairs is sufficiently well described by Gaussian distribution law with small non-linear corrections.

However, it would be more efficient to create an orthonormal system localized in time domain, in particular – to represent initial series as wavelet expansion.

Let us introduce the set of analyzing functions - wavelets $\psi_{i\tau}(t) = 2^{-i/2}\psi(2^{-i}t - \tau)$, meeting conditions $\psi(t) \subset L^1 \cap L^2, \int \psi(t)dt = 0$ and forming an orthonormal basis for $L^2(R)$ with compact carrier. Minor value of index $i$ corresponds to the low-frequency component of the analyzed function $X(t)$, while the major one – to the high-frequency component.

Let us represent the analyzed series as linear combination of functions $\psi_{i\tau}(t)$ with coefficients $Y_i(t) = \int X(t)\psi_{i\tau}(t)dt$. Let us denote $\mathbf{Y}(t) = (Y_1(t),..Y_K(t))$, where - $K$ maximal order of the wavelet transformation.

If initial sequence is represented as a discrete series, then the wavelet expansion coefficients are only linear combination of corresponding values of the analyzed function within the carrier's length [ 4 ]:

$$Y_{ik} = \sum_p h_p Y_{i-1,2k+p}, Y_{0k} = X_k$$

Coefficients of this linear combination $h_k$ for known wavelets and spline-wavelets have been calculated in advance with needed accuracy. Advantage of this description is orthogonality in space and localization in time, that is expansion coefficients describe the system's behavior in the description point's vicinity at various time scales, and thereby they principally differ from conventional Fourier transformation. Other advantage of the wavelet expansion is possibility to implement them when creating fast computer algorithms. The most simple wavelet – is Haar wavelet. Within carrier (interval) "1" it is defined as $\psi_{10}(t) = (1/\sqrt{2})$ $0 < t < 1/2; \psi_{10}(t) = -(1/\sqrt{2})$ $1/2 < t < 1$.

It is easy to observe that up to the sign's accuracy the expansion coefficient of the Haar wavelet – is the analyzed value's change rate smoothed over the carrier. On the other hand, by direct substitution one can become sure that it is difference between two smoothings of direct sequence with unit and double steps.

Hence, Haar wavelet expansion coefficients – are amplitudes of the various-length waves of the studied value's change rate in given point (at given time). For other wavelets the expansion coefficients have no such evident interpretation, though their orthogonality and localization in time are preserved.

Question, which is principal for further understanding, is origin of dependence $X(t)$, or $X_n$ in discrete representation. This is either random process, or realization of process of the so-called deterministic chaos. Deterministic chaos or chaotic regimes in finite-dimensional systems or discrete mappings – is comparatively new area of mathematics, ascending to famous results of Feigenbaum for simple mappings and to so-called chaos in the system of Lorenz equations.

In particular, Feigenbaum managed to show that for simple unidimensional mapping $X_{n+1} = \lambda x_n(1-x_n)$ at definite values of parameter $\lambda$, for instance, at $\lambda \approx 3.618...$ or $\lambda \approx -1.62...$ the mapping becomes chaotic, that is system's behavior becomes unpredictable, and unlimitedly grows when returning to infinitesimal neighborhood of any point. Attracting manifold, which in this case appears in phase space of such system, is called strange attractor.

For external observer the system's behavior becomes chaotic, and realizations are perceived as random ones. Later on the chaotic systems have been created for two-dimensional mappings, such as baker's mapping (transformation), horseshoe mapping, etc. Models of deterministic chaos have been created, for instance – transition to chaos through cascade of the period doubling bifurcations, as well as a theorem has been proved about auxiliary mappings, which theorem plays role of necessary condition for possibility of appearance of the strange attractor in two-dimensional mappings.

Important condition of possibility of chaotic regimes is presence of immovable hyperbolic points, that is condition of that the modulus of the real part of one of the eigenvalues of the mapping linearized in the immovable point is greater that 1.

Similar results exist for the model composed of the system of autonomous differential equipment. The only limitation is that this system should have dimensionality more than 2, and at least one hyperbolic singular point should exist in it, that is immovable point, in which real part of one of their eigenvalues is positive.

Further analysis is possible with use of the Poincare map, that is consideration, instead of the initial continuous system, of the lower-dimensionality system obtained by sequential intersection of phase trajectories with two-dimensional surface – Poincare sections. Sequences of the intersection points, generally speaking, are connected by some interrelations, which may be analyzed using methods of creation of chaotic structures for the mappings.

Attempts to find discrete mappings (small-mode chaos) for various financial sequences (from Forex currency pair quotations to indices) up to date were not successful. In any case, in the studies conducted by author of this paper, for great number of initial sequences up to dimensionality $Q=5$, that is mappings of type $X_{n+1} = f(X_n, X_{n-1}, .... X_{n-5})$ , no necessary preconditions appear for emersion of chaotic regimes or regimes close to them (for instance, transient chaos of Popot-Manneville type).

To say it more exactly, such preconditions almost do not appear. Possibilities of chaotic surges for various time frames have been observed in the vicinity of the financial crisis commencement in the mid-September 2008 for indices DJ, S@P500, and some others. Still there is no unambiguous answer to possibility of use of this criterion for forecasting of stock market downfalls some hours prior such downfalls happen.

Hence, almost always, excluding neighborhood of essential downfalls, oscillations of the financial sequences, or of some mappings of these sequences on that or other orthogonal spaces, are most probably described by stochastic models, whose construction is dealt with in the next Section of this paper.

### 3. Langevin and Ito equations, path integrals. Possibility of equilibrium and nonstationarity

The simplest stochastic equation possessing essential generality and possibility of rational interpretation and analysis – is the Langevin equation, given, for instance in the Ito form

$$dx(t) = F(x)dt + G(x)d\omega(t)$$

Here $x(t)$ - is some continuous or discrete variable – initial financial sequence or its mapping, for instance, first coefficient of $Y_1(t)$ wavelet – expansion at time $t$.

Value $d\omega(t)$ realizes random, delta-correlated Gaussian process with zero mean values $\langle d\omega \rangle = 0, \langle d\omega(t)d\omega(t_1) \rangle = \delta(t-t)$.

For random proves it is possible to determine conditional probability $f(x|x_0)$ meeting Fokker-Planck equation

$$\frac{\partial f(x|x_0)}{\partial t} = \frac{\partial}{\partial x}(F(x)f(x|x_0)) + \frac{1}{2}\frac{\partial}{\partial x}(\frac{\partial}{\partial x}G^2(x)f(x|x_0))$$

Solution of the last equation may be presented as a path integral, for instance, by means of transformations described in [1]. In particular, when we assume linearity of the functions

$F(x) = \lambda_1 x, G(x) = \lambda_2 x$, continual integral may be computed explicitly, and solution coincides with the Black-Scholes formula received, naturally, from other considerations.

Stationary solutions of this equation are of special interest. In particular, stationary solution for unidimensional case is received by elementary integration. Stationary distribution with condition $x(0) = x_0$ looks like

$$f(x) = Z^{-1} \exp(\int_{x_0}^{x} \frac{F(x)}{2G^2(x)} dx)$$, (1)

where $Z$ - is normalizing factor.

In case of multidimensional problem the expression for conditional probability may also be represented as a path integral, however, stationary solution requires additional conditions, the so-called compatibility conditions. When these conditions are fulfilled, the stationary solution is obtained, like in the unidimensional case, by single-time integration.

The same problem also appears for the so-called non-Markovian processes, when the equation and Ito increment of value $dx_n$ depends not only on values of the variable at time $t$, but also on its values at previous times. In discrete representation the relevant equation acquires the following form $dx_n = F(x_n, x_{n-1}...)dt + G(x_n)d\omega$.

In continuous case we have an integral-differential equation with lagged cores. Their analysis is not so simple in our times.

Another method to obtain the non-Markovian effects in the equation and the Ito – is to use wavelet-expansion coefficients in the next (double) scale $dY_{n1} = F(Y_{n1}, Y_{n2})dt + G(Y_{n1})d\omega$. In this case it is possible to create a system of interlocking equations allowing for various-scale time effects. It is similar system that has been analyzed in [2].

Having cut the system at a particular wavelet (scale), it is possible to express, by known preceding points, the increment value $dY_{n1}$, that is to implement forecast of the system's behavior. Having integrated (1) for the Markovian approximation, it is possible to determine stationary distribution.

Thereafter, there is nothing to do but to create description of the drift $F(Y_{n1}, Y_{n2})$ and diffusion $G(Y_{n1})$ summands of the Ito equation by available observations, that is to solve a conventional optimization problem $E(Q(x, \lambda)) \to \min$, where $x = (Y_{n1}, Y_{n2})$ - the set of variables observed at various times, symbol $E$ denotes averaging over observations, while minimization is performed over the set of parameters $\lambda = (\lambda_1, \lambda_2....)$ of series expansion of value $F, G$ by an appropriate set of orthogonal functions (in our case – Hermite polynomials), $Q(x, \lambda)$ - certain norm, for instance, square one $Q(Y_{n1}, Y_{n2}) = (dY_{n1} - F(Y_{n1}, Y_{n2}))^2$.

Actually it is a standard problem of filtration, forecast, minimization of risk, etc. it is exactly how the problem was being solved for the first variant of the model [2].

If density of distribution is known, then the averaging is changed by integration, and nothing is left from the physical essence of the problem. However, this density is actually unknown, moreover, the analyzed process may appear to be non-stationary, or to say it more properly, a "slow time" may exist, in whose scales the parameters of the analyzed statistics are varying.

Therefore we subsequently will use a method based on the so-called adaptation procedure – adjustment of the model parameters in the course of observations of the object's evolution (Robbins-Monro procedure [5]).

Namely, for two time moments $n, n+1$ we have created procedure $\lambda_k(n+1) = \lambda_k(n) + \beta \phi(\lambda(n), x(n))$. This procedure converges by probability to solution of the equation $E(\phi(x, \lambda)) = 0$, if we assume $\phi(x, \lambda) = \nabla Q(x, \lambda)$.

Maximum simplification of this condition means that at each step of observation of the system (at each time moment) the expansion coefficients in description of drift and diffusion components of the Ito equation are changing so that in the average to minimize the difference between the observations and the descriptions.

This algorithm allows to rather efficiently create description for current state of the system, and to allow for slow alteration of the model's constants in the course of its evolution. Let us assume that $T$ is characteristic slow time scale. If the description is created, then in Markovian approximation it is possible то integrate (1) and to obtain stationary distribution $f_s(Y_1)$. Naturally, for stationary distribution we can omit to write index $n$ in the expression for the distribution density.

If there are two stationary distributions obtained by time shift $t_0 \approx T$, then using Kolmogorov non-parametric statistical criteria it is possible to evaluate significance of their difference with risk level $\alpha$.

In other words, synthesized stationary distributions $f_s'(Y_1), f_s'(Y_1)$ at shift $t \to t+T$ with probability $p = 1 - \alpha$ are deemed to be indistinguishable, if $\max |K(Y_{n1}) - K'(Y_{m1})| < k(\alpha)/\sqrt{N}$, где $K, K'$ cumulative probabilities for this state, and for state obtained by shift $T$.

Points $m, n$ correspond to time moments $t$ и $t+T$. Constant $k(\alpha) \approx 1$ depends on risk level $\alpha$, $N$ - is number of experimental points (measurements), over which the approximation of stationary distribution have been created.

Therefore, in the most simple case the optimum purchase time (entry into the long position) may be represented as $t = \inf(-dY_1 > 0, P_s > 1 - \alpha_1)$, while the optimum sales time (entry into the short position) $t = \inf(-dY_1 < 0, P_s < \alpha_1)$, where $P_s = \int_{-\infty}^{0} f_s(y) dy$.

Here we have used a composite criterion: simultaneous fulfillment of the condition of positive increment of the averaged rate of change of the currency pair quotation (sign minus is due to the Haar wavelet) and excess of probability of the positive trend - value $1 - \alpha_1$.

Alternative variant – is to use convolution $f_s(z) = \int_{-\infty}^{+\infty} f_s(y_1) f_{s,T}(y_1 + z) dy_1$. In this case conditions $P_s > 1 - \alpha_1, P_s < \alpha_1$ acquire sense of the time of resale and repurchase [2].

In the both cases it is necessary to exclude statistically insignificant differences between the stationary distributions at time shifts $t \to t+T$, what is done by use of the Kolmogorov-Smirnov non-parametric test.

Criteria connected with sign $dY_1$ will be subsequently called elementary dynamic ones, while that with probabilistic assessments – will be called elementary statistical criteria.

## 4. Structures from chaos. Detailed description of the algorithm

Principal scheme of the algorithm is given in Figure 1, while its brief description is given in [ 3 ].

Generators of the elementary solutions are denoted as 1 and 2, and correspond to statistical and dynamical criteria of [3]. Total number of blocks is unlimited; as such blocks one can use standard elementary models – method of moving averages (MACD), Bollinger bands (BB), relative market strength index, etc. Apart from the current value of the currency pair quotations, as input information for decision making it is possible to use stock market indices, futures, prices for gold, etc.

The elementary solution blocks (1-3) generate, on the grounds of their own logics, the conditions of entry to and exit from the short and long positions, for instance, in the variant given in Section 3 hereof. Let us assume that at exit from $i$ such blocks for $k$ currency pairs the variable $u_{ik}$ becomes +1 for entry into the long position and -1 entry into the short position.

Complexity and instability of the initial process is such that in advance it is impossible to say which one of the elementary criteria is efficient. Therefore, at the next stage we use elements of Boolean algebra: sum of entries is converted by means of relationship $U_k = th(\alpha_1 u_{1k} + \alpha_2 u_{2k} + ...)$ , while weight of the elementary solutions $\alpha_k$ is selected so that to minimize norm of differences between the sign of quotation change $A_k = +-1$ and decision made on the grounds of dynamical and statistical criteria, or other criteria $(A_k - U_k)^2 \to \min$. Here we once again use the Robbins-Monro procedure of dynamical adjustment - Block 4.

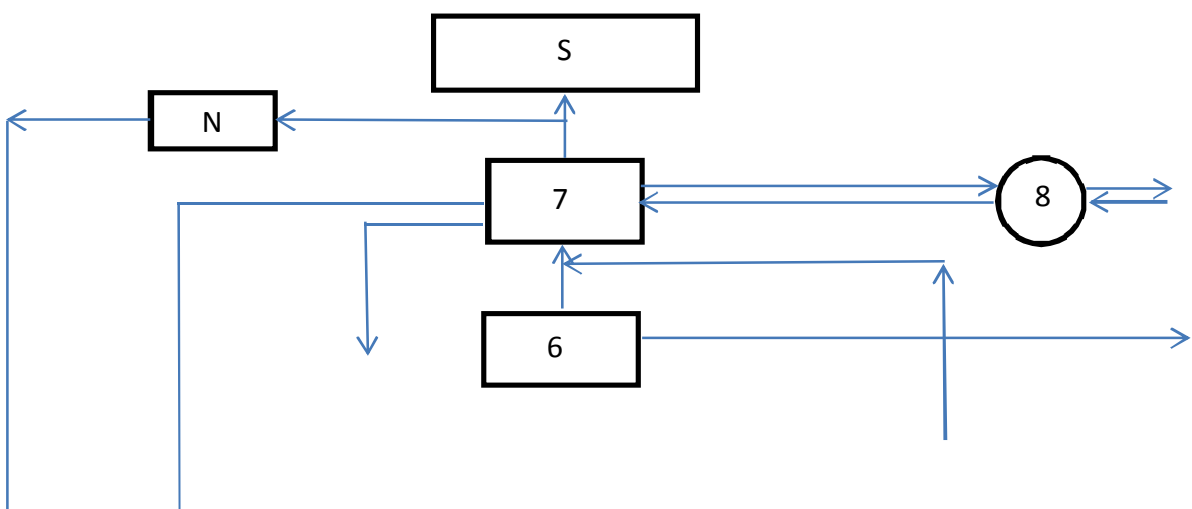

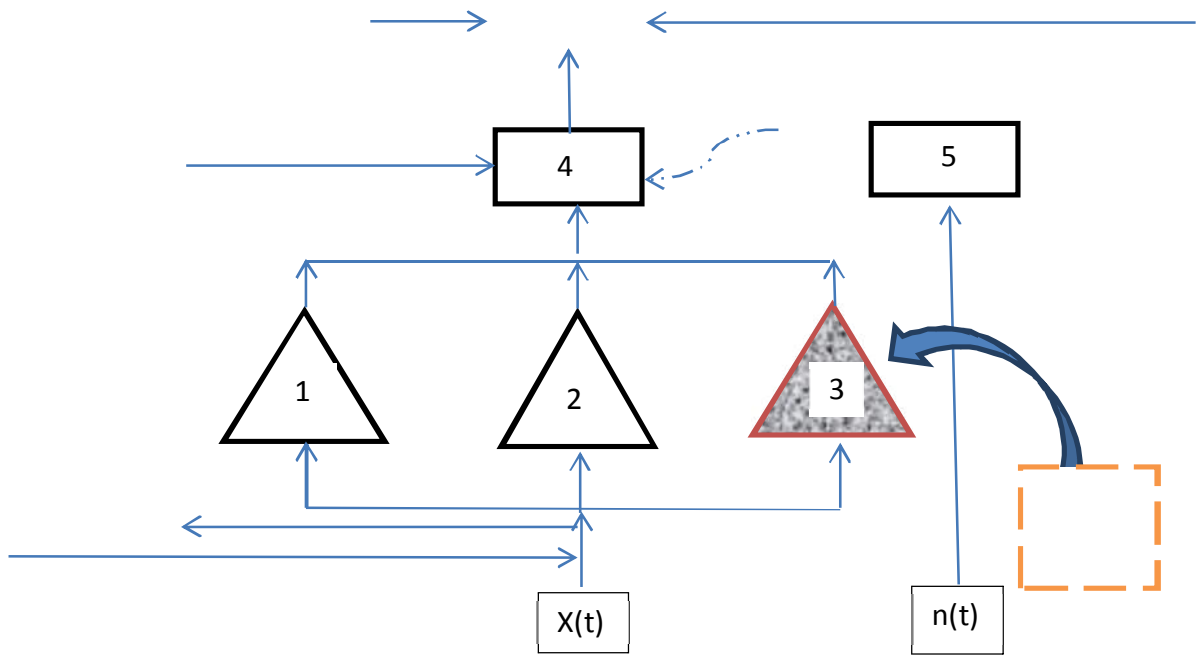

Figure 1. Principal scheme of the Structure from Chaos algorithm of the NSW model.

1,2.- generators of elementary solutions (model NSW); 3 - block of self-similarity assembly; 4 - block of dynamic optimization and generation of structures; 5 - block of package optimization; 6 - block of collective decision making; 7 - block of generation of closely-coupled self-assembly; 8 - block of generation of loosely-coupled self-assembly; N - block of vertical self-assembly; S – central trade server; X(t), n(t) – current quotation and share of the currency pair in the financial tools package.

If the optimization goal is attained – that is direction of change is correctly forecasted, then output of the block 4 – is value $U_k$, otherwise (signs are different) value $U_k$ is multiplied by compensation factor (usually 0.5), which mitigates effect of the initial decision made for given currency pair on making of decisions for other currency pairs. This variant of interaction inside the multicurrency advisor was called closely-coupled horizontal self-assembly.

Idea of the closely-coupled horizontal self-assembly is based on the fact that even with use of the Robbins-Monro adjustment procedure the available information is not always sufficient for efficient restoration of the movement equation and for creation of quasistationary distribution; in other words, rate of change of the description statistics in the slow time scale is high enough for their due-quality and significant description.

In other words, in the majority of cases we are working at the edge of possibility to separate the processes into "fast" and "slow" ones – the movement equation and the quasistationary distributions. By the way, it is from this situation that inefficiency of the neural net algorithms is resulting: adjustment of the net (pattern recognition) requires learning sequence's length to be much more greater than time scales interesting for us (of slow time) – and obtained results arte trivial and can not be used as basis of an efficient advisor.

This problem may be solved by participation in solving of the question about entering-exiting into/from the trade positions for given currency pair of information on other currency pairs. To this end we compute paired correlation coefficients of quotations of all currency pairs $\lambda_{kp}$ and determine integral characteristics $S_k = \lambda_{k1}U_1 + \lambda_{k2}U_2 + ...$ - as sum of the announced decisions for individual pairs with allowance for their intercorrelation. We allow only for correlations with significance level $\alpha = 0.05$.

If two currency pairs have positive correlation coefficient, than decision made in block 4 to enter the short position for one currency pair will increase probability of similar decision for other currency pair, and vice versa in case of negative correlation.

Hence, the decision making blocks may be in various statuses/states – in the state of active trade (making decision and interaction with the trade server - block S), in semi-active state – discussion of the decisions and interchange of information with other blocks, and in passive state – disconnected from the decision making and discussion processes. Transition between these stages for each currency pair is determined according to quality of forecasts of the future quotation changes (block 7). If share of correctly forecasted directions of the exchange rate alterations exceeds 0.5, then relevant control signal arrives to the trade server S (active state), if this share is from 0.25 to 0.5 – then the state is semi-active, and if it is less than 0.25 – then the state is passive.

Another problem solved in the proposed algorithm – is the problem of redistribution of the assets share invested in short or long positions for various currency pairs, that is compromise between maximum profitability and minimum risk - block 8.

In this block the so-called mixed strategy is simulated on the basis of the completed transaction statistics. Similar description of this algorithm is described in [2]. If $n_i$ is share of assets invested in given package for a currency pair $i$, $x_i$ - is average profitability for closed positions, $R_{ij}$ - paired correlation function of profitability, then we are to solve the problem of maximization of the value $J = n_i x_i + \beta n_i n_j R_{ij} \to \max$, where $\beta$ is the compromise parameter between maximum profitability and minimum risk. Usually selected value is $\beta = 0.25$, though value $\beta$ adjustable parameter of the algorithm. This block is called loosely-coupled self-assembly block.

Origin of notions of the closely- and loosely-coupled self-assembly is connected with characteristic time scales for recalculation of the problem's parameters: for the closely-coupled self the parameters are being recalculated for each tick, that is up to 100 times a minute depending on market activity, while the loosely-coupled self-assembly carries out the recalculation depending on number of closed transactions – on an average once in 1..2 hours.

Finally, the algorithm provides for two other methods to influence the decision making, firstly, it is possibility to use all the structure as a whole as the elementary decisions block, but in another time scale (with other time frame) – block 3, secondly, it is possibility of the so-called vertical self-assembly: obtained decision (in particular, entry into or exit from the short or long position) at given time moment may be used as input variables together with the current quotations - block N. In a number of cases it allows to make strategic decisions, however, requirements to performance of the computer system drastically increase.

Apart from this, the algorithm was built so as to provide for automatic setting of the Take Profit position depending on value the root-mean-square deviation of the current exchange rate from the smoothed one in the "slow" time scale; similar method was used for setting of the Tralling Stop level, and, additionally, it was provided for compulsory closing of the position, floating loss wherefrom was exceeding a preset share of ready assets.

For each currency pair the algorithm allowed to open up to two positions in any directions, while opening of the second position was permitted in case of presence of unclosed first position, provided currently it is loss-making, only after expiry of definite time period.

## 5. Real-time embodiment of the algorithm. Probability of profitable trade

Proposed algorithm has been tested in automatic mode in real time for simultaneous trade of eight currency pairs. The self-similar and vertical self-assembly were not used. Results for various time intervals are given in Table 1. Time frame was 1 minute, base for the Haar wavelet determination $p_1 = 5$, risk levels $\alpha = \alpha_1 = 0.15$. Initial deposit was $5000.

Table 1. Algorithm's operation efficiency : year 2011, 8 currency pairs.

| Period | Number of transactions | Profit for closed positions $ | Floating profit (loss),$ | Profit after closing of open positions ,$ | Probability of profitable trade |
|---|---|---|---|---|---|
| 09.06-04.08 | 390 | 20568.56 | -4750.19 | 15818.37 | 0.87 |
| 10.08-13.09 | 205 | 10287.78 | -1044.86 | 9242.92 | 0.78 |
| 19.08-26.09 | 428 | 7220.22 | -1512.03 | 5708.19 | 0.67 |
| 3.10-11.10 | 49 | 1479.81 | -912.77 | 567.04 | 0.81 |

Probability of profitable trade was evaluated on the grounds of the completed transaction statistics. Transaction connected with emergency closing of the position, as well as compulsory closed loss-making positions as of the end of analyzed period were both deemed to be loss-making transactions. Further we calculated average profitability of transaction (normalized to current balance) and its root-mean-square deviation. Empirical histogram was compared with theoretical normal distribution at certain risk level using Kolmogorov criterion (risk was assumed to be 0.05).

Then we calculated the loss-free trade probability and, generally speaking, probability of trade at any predefined level of profit. Relevant data is given in Table 1.

The algorithm has generally shown good result. All analyzed periods appeared to be profitable, weekly profit was in the range of 0.2-0.35. The most unsuccessful period 19.08-26.09, which was accompanied by three emergency closings of greatly loss-making short positions eur/aud and aud/usd, finished, however, as successful (Annex 1).

### 6. When the algorithm makes errors. Whether the model will be serviceable in conditions of global financial catastrophe?

As it has been said above, process of formation of the financial sequences is most probably random (Markovian or non-Markovian process). Therefore any forecasts obtained on the grounds of these models have probabilistic nature. Good algorithm differs from the bad one only in that the probability of correct forecast is somewhat higher than of the incorrect one with not too big dispersion.

Suggested algorithm in a number of cases makes erroneous forecasts due to the following cause.

Entry into short or long positions is performed correctly, then for the given currency pair the system is balancing on small profits and losses insufficient for closing of the position. If after that an abrupt growth or drop of quotations occur, then emergency closing of the position is performed with significant loss. Typical instance are short positions eur/aud and aud/usd opened on 12.09.2011 and 20.09.2011 and closed by emergency criterion.

To avoid this, the algorithm should be complemented by closing condition on the Stop Loss level, whose value depends on time interval from the moment of position opening. Time interval and dependence of closing level for each currency pair – is a typical optimization problem solvable in our times.

World financial system in the nearest future is to suffer shocks surpassing the 2008 events. No "quantitative mitigations" Q2,Q3…. are able to prevent it. One of the models of this financial catastrophe is given in [6].

With this, the world financial system's behavior in the catastrophe's neighborhood inevitably will feature pronounced volatility of the Forex market, namely, expressly volatile market is the most suitable time for financial speculations with use of automatic trade systems. Testing of the algorithm for expressly volatile sequences (common financial series modulated by periodical function with addition of random component), has shown that here profitability and stability of the algorithm's operation is higher than for the initial sequences.

### 7. Perspectives of the model development

Model and algorithm are currently an "erector set" composed of elementary blocks and allowing to build rather complicated structures. These blocks are interconnected by rather flexible links allowing the structure to adjust to changing external conditions.

It is obviously interesting to complement the model with a block of analysis of text information – reports of information and rating agencies. Logic of operation of such block is obvious. It will trace and range critical terms, such as "act of terrorism", "general strike", "interest rate decline", "armed conflict", etc. Should it detect such event, an emergency exit is performed, or some other position is entered. For instance, act of terrorism in country "A" requires immediate closing of long position with currency of this country in the numerator, and, probably, opening of a short position, etc.

With this, the author deems it necessary to note certain paradox consisting in that the more factors are allowed for, the worse the model becomes. Use of small number of variables testifies about correctness of the model, while their great number testifies to contrary. Function of great number of factors (variables) is almost constant in the vicinity of maximum, which slightly exceeds average value. It is why errors of input data, incorrect interpretation, or excessive variables bring to nought the accurate solution of the problem.

Annotation

This paper deals with detailed description of the multicurrency trade algorithm based on NSW model with use of flexible algorithm capable of adaptation to market conditions.

Abstract

Flexible algorithm of multicurrency trade on Forex market has been built on the grounds of non-linear stochastic wavelets (NSW) model. Probability of the loss-free trade has been evaluated. Results of the algorithm's real-time testing and issues of the algorithm's development are discussed.

**Account:** 3380748  **Name:** Алексей Авдеенко  **Currency:** USD  **2011 September 26, 20:57**

**Closed Transactions:**

| Ticket | Open Time | Type | Size | Item | Price | S / L | T / P | Close Time | Price | Commission | Taxes | Swap | Profit |
|---|---|---|---|---|---|---|---|---|---|---|---|---|---|
| 129621195 | 2011.08.19 19:52 | sell | 0.10 | usdjpy | 76.370 | 0.000 | 0.000 | 2011.09.20 12:52 | 76.452 | 0.00 | 0.00 | -2.60 | -10.73 |
| 129619558 | 2011.08.19 19:32 | sell | 0.95 | gbpusd | 1.65503 | 1.65496 | 1.65083 | 2011.08.19 19:46 | 1.65446 | 0.00 | 0.00 | 0.00 | 54.15 |
| 128984743 | 2011.08.16 21:42 | buy | 0.23 | gbpusd | 1.64629 | 1.64633 | 1.65057 | 2011.08.16 22:50 | 1.64633 | 0.00 | 0.00 | 0.00 | 0.92 |
| 129122529 | 2011.08.17 14:03 | buy | 0.19 | gbpusd | 1.64597 | 1.55574 | 1.65024 | 2011.08.17 14:57 | 1.65024 | 0.00 | 0.00 | 0.00 | 81.13 |
| 128994126 | 2011.08.16 23:17 | sell | 0.15 | gbpusd | 1.64545 | 1.64422 | 1.64115 | 2011.08.17 08:24 | 1.64371 | 0.00 | 0.00 | -0.60 | 26.10 |
| 129070531 | 2011.08.17 10:27 | sell | 0.23 | gbpusd | 1.64358 | 1.64234 | 1.63929 | 2011.08.17 11:24 | 1.64169 | 0.00 | 0.00 | 0.00 | 43.47 |
| 128931280 | 2011.08.16 17:51 | sell | 1.98 | gbpusd | 1.64140 | 1.64065 | 1.63718 | 2011.08.16 19:30 | 1.64065 | 0.00 | 0.00 | 0.00 | 148.50 |
| 129089632 | 2011.08.17 11:43 | buy | 0.22 | gbpusd | 1.63753 | 1.63791 | 1.64182 | 2011.08.17 11:45 | 1.63791 | 0.00 | 0.00 | 0.00 | 8.36 |
| 128362767 | 2011.08.11 17:49 | sell | 0.44 | gbpusd | 1.62306 | 1.71326 | 1.61876 | 2011.08.11 17:54 | 1.62273 | 0.00 | 0.00 | 0.00 | 14.52 |
| 128363223 | 2011.08.11 17:50 | sell | 0.20 | gbpusd | 1.62287 | 1.62259 | 1.61860 | 2011.08.11 18:00 | 1.62209 | 0.00 | 0.00 | 0.00 | 15.60 |
| 128367487 | 2011.08.11 18:04 | buy | 0.12 | gbpusd | 1.62266 | 1.53247 | 1.62697 | 2011.08.12 11:49 | 1.62697 | 0.00 | 0.00 | -0.14 | 51.72 |
| 128367480 | 2011.08.11 18:04 | sell | 0.12 | gbpusd | 1.62247 | 1.71266 | 1.61816 | 2011.08.11 18:23 | 1.62076 | 0.00 | 0.00 | 0.00 | 20.52 |
| 128366925 | 2011.08.11 18:02 | sell | 0.22 | gbpusd | 1.62233 | 1.71254 | 1.61804 | 2011.08.11 18:23 | 1.62077 | 0.00 | 0.00 | 0.00 | 34.32 |
| 128419454 | 2011.08.11 22:57 | sell | 0.19 | gbpusd | 1.62212 | 1.62015 | 1.61783 | 2011.08.12 08:39 | 1.62015 | 0.00 | 0.00 | -0.68 | 37.43 |
| 128365401 | 2011.08.11 18:00 | buy | 0.16 | gbpusd | 1.62212 | 1.62232 | 1.62644 | 2011.08.11 18:02 | 1.62232 | 0.00 | 0.00 | 0.00 | 3.20 |
| 128376689 | 2011.08.11 18:47 | buy | 0.38 | gbpusd | 1.62125 | 1.62214 | 1.62554 | 2011.08.11 19:00 | 1.62214 | 0.00 | 0.00 | 0.00 | 33.82 |
| 128376687 | 2011.08.11 18:47 | sell | 0.38 | gbpusd | 1.62104 | 1.71125 | 1.61675 | 2011.08.11 20:05 | 1.62059 | 0.00 | 0.00 | 0.00 | 17.10 |
| 128371847 | 2011.08.11 18:23 | buy | 0.30 | gbpusd | 1.62094 | 1.62118 | 1.62522 | 2011.08.11 18:46 | 1.62118 | 0.00 | 0.00 | 0.00 | 7.20 |
| 128394049 | 2011.08.11 20:17 | sell | 0.25 | gbpusd | 1.62078 | 1.62065 | 1.61643 | 2011.08.11 20:27 | 1.62014 | 0.00 | 0.00 | 0.00 | 16.00 |
| 128391656 | 2011.08.11 20:05 | buy | 0.31 | gbpusd | 1.62060 | 1.62064 | 1.62492 | 2011.08.11 20:17 | 1.62064 | 0.00 | 0.00 | 0.00 | 1.24 |
| 128396165 | 2011.08.11 20:27 | buy | 0.24 | gbpusd | 1.62009 | 1.52991 | 1.62441 | 2011.08.11 21:13 | 1.62077 | 0.00 | 0.00 | 0.00 | 16.32 |
| 128469324 | 2011.08.12 08:39 | sell | 0.23 | gbpusd | 1.61990 | 1.61973 | 1.61565 | 2011.08.12 10:09 | 1.61973 | 0.00 | 0.00 | 0.00 | 3.91 |
| 128483755 | 2011.08.12 10:09 | sell | 0.10 | gbpusd | 1.61970 | 1.61841 | 1.61538 | 2011.08.12 10:13 | 1.61841 | 0.00 | 0.00 | 0.00 | 12.90 |
| 128271217 | 2011.08.11 11:39 | buy | 0.20 | gbpusd | 1.61923 | 1.62255 | 1.62341 | 2011.08.11 17:49 | 1.62302 | 0.00 | 0.00 | 0.00 | 75.80 |
| 128271214 | 2011.08.11 11:39 | sell | 0.20 | gbpusd | 1.61891 | 1.61855 | 1.61473 | 2011.08.11 11:49 | 1.61802 | 0.00 | 0.00 | 0.00 | 17.80 |
| 128271180 | 2011.08.11 11:39 | sell | 0.34 | gbpusd | 1.61886 | 1.70907 | 1.61457 | 2011.08.11 11:44 | 1.61833 | 0.00 | 0.00 | 0.00 | 18.02 |
| 128126556 | 2011.08.10 16:49 | sell | 0.13 | gbpusd | 1.61876 | 1.61781 | 1.61460 | 2011.08.10 16:58 | 1.61728 | 0.00 | 0.00 | 0.00 | 19.24 |
| 128488074 | 2011.08.12 10:20 | sell | 0.19 | gbpusd | 1.61860 | 1.61731 | 1.61430 | 2011.08.12 10:37 | 1.61731 | 0.00 | 0.00 | 0.00 | 24.51 |
| 128261434 | 2011.08.11 10:42 | buy | 0.20 | gbpusd | 1.61851 | 1.61876 | 1.62280 | 2011.08.11 11:39 | 1.61876 | 0.00 | 0.00 | 0.00 | 5.00 |
| 128125779 | 2011.08.10 16:45 | buy | 0.15 | gbpusd | 1.61837 | 1.61878 | 1.62253 | 2011.08.10 16:49 | 1.61878 | 0.00 | 0.00 | 0.00 | 6.15 |
| 128273817 | 2011.08.11 11:50 | buy | 0.35 | gbpusd | 1.61830 | 1.62198 | 1.62259 | 2011.08.11 17:47 | 1.62259 | 0.00 | 0.00 | 0.00 | 150.15 |
| 128262652 | 2011.08.11 10:47 | buy | 0.25 | gbpusd | 1.61796 | 1.61847 | 1.62223 | 2011.08.11 11:37 | 1.61847 | 0.00 | 0.00 | 0.00 | 12.75 |
| 128132367 | 2011.08.10 17:20 | sell | 0.13 | gbpusd | 1.61763 | 1.61733 | 1.61348 | 2011.08.10 17:24 | 1.61733 | 0.00 | 0.00 | 0.00 | 3.90 |
| 128172176 | 2011.08.10 21:13 | sell | 0.35 | gbpusd | 1.61754 | 1.70787 | 1.61337 | 2011.08.10 21:16 | 1.61702 | 0.00 | 0.00 | 0.00 | 18.20 |
| 128134120 | 2011.08.10 17:28 | sell | 0.13 | gbpusd | 1.61743 | 1.61651 | 1.61327 | 2011.08.10 17:52 | 1.61582 | 0.00 | 0.00 | 0.00 | 20.93 |
| 128129007 | 2011.08.10 16:58 | buy | 0.13 | gbpusd | 1.61743 | 1.61761 | 1.62160 | 2011.08.10 17:20 | 1.61761 | 0.00 | 0.00 | 0.00 | 2.34 |
| 128133186 | 2011.08.10 17:24 | buy | 0.13 | gbpusd | 1.61736 | 1.61758 | 1.62154 | 2011.08.10 17:28 | 1.61758 | 0.00 | 0.00 | 0.00 | 2.86 |
| 128254361 | 2011.08.11 10:20 | buy | 0.24 | gbpusd | 1.61725 | 1.61825 | 1.62155 | 2011.08.11 10:46 | 1.61825 | 0.00 | 0.00 | 0.00 | 24.00 |
| 128492710 | 2011.08.12 10:37 | sell | 0.21 | gbpusd | 1.61716 | 1.70735 | 1.61285 | 2011.09.05 10:32 | 1.61285 | 0.00 | 0.00 | -16.46 | 90.51 |
| 128254358 | 2011.08.11 10:19 | sell | 0.24 | gbpusd | 1.61705 | 1.70725 | 1.61275 | 2011.08.11 10:25 | 1.61600 | 0.00 | 0.00 | 0.00 | 25.20 |
| 128174911 | 2011.08.10 21:29 | sell | 0.41 | gbpusd | 1.61705 | 1.70738 | 1.61288 | 2011.08.10 21:38 | 1.61668 | 0.00 | 0.00 | 0.00 | 15.17 |
| 128172937 | 2011.08.10 21:17 | buy | 0.29 | gbpusd | 1.61704 | 1.61711 | 1.62122 | 2011.08.10 21:29 | 1.61711 | 0.00 | 0.00 | 0.00 | 2.03 |
| 128176231 | 2011.08.10 21:44 | buy | 0.10 | gbpusd | 1.61677 | 1.61707 | 1.62094 | 2011.08.11 10:19 | 1.61707 | 0.00 | 0.00 | -0.27 | 3.00 |
| 128173061 | 2011.08.10 21:17 | buy | 0.10 | gbpusd | 1.61676 | 1.61708 | 1.62094 | 2011.08.10 21:29 | 1.61708 | 0.00 | 0.00 | 0.00 | 3.20 |
| 128175815 | 2011.08.10 21:38 | buy | 0.40 | gbpusd | 1.61667 | 1.61708 | 1.62084 | 2011.08.11 10:19 | 1.61708 | 0.00 | 0.00 | -1.08 | 16.40 |
| 128168857 | 2011.08.10 20:58 | buy | 0.24 | gbpusd | 1.61600 | 1.61727 | 1.62026 | 2011.08.10 21:13 | 1.61760 | 0.00 | 0.00 | 0.00 | 38.40 |

| ID | Open Time | Type | Size | Symbol | Price | SL | TP | Close Time | Close Price | Commission | Taxes | Swap | Profit |
|---|---|---|---|---|---|---|---|---|---|---|---|---|---|
| 128139996 | 2011.08.10 17:53 | buy | 0.12 | gbpusd | 1.61580 | 1.61731 | 1.61999 | 2011.08.10 21:13 | 1.61731 | 0.00 | 0.00 | 0.00 | 18.12 |
| 132773575 | 2011.09.12 15:11 | sell | 0.56 | gbpusd | 1.58697 | 1.58644 | 1.58265 | 2011.09.12 16:08 | 1.58644 | 0.00 | 0.00 | 0.00 | 29.68 |
| 132767320 | 2011.09.12 14:50 | buy | 1.30 | gbpusd | 1.58679 | 1.58683 | 1.59111 | 2011.09.12 15:10 | 1.58683 | 0.00 | 0.00 | 0.00 | 5.20 |
| 133589314 | 2011.09.15 17:12 | sell | 0.24 | gbpusd | 1.58637 | 1.58595 | 1.58205 | 2011.09.15 17:16 | 1.58559 | 0.00 | 0.00 | 0.00 | 18.72 |
| 132789946 | 2011.09.12 16:15 | sell | 1.24 | gbpusd | 1.58586 | 1.58479 | 1.58155 | 2011.09.12 16:27 | 1.58479 | 0.00 | 0.00 | 0.00 | 132.68 |
| 133590419 | 2011.09.15 17:15 | sell | 0.79 | gbpusd | 1.58562 | 1.58531 | 1.58131 | 2011.09.15 17:21 | 1.58531 | 0.00 | 0.00 | 0.00 | 24.49 |
| 133591893 | 2011.09.15 17:21 | sell | 1.39 | gbpusd | 1.58512 | 1.58346 | 1.58081 | 2011.09.15 17:30 | 1.58346 | 0.00 | 0.00 | 0.00 | 230.74 |
| 132802019 | 2011.09.12 16:56 | sell | 0.57 | gbpusd | 1.58410 | 1.58395 | 1.57977 | 2011.09.12 17:36 | 1.58395 | 0.00 | 0.00 | 0.00 | 8.55 |
| 132812579 | 2011.09.12 17:36 | sell | 0.78 | gbpusd | 1.58382 | 1.58269 | 1.57952 | 2011.09.12 17:44 | 1.58269 | 0.00 | 0.00 | 0.00 | 88.14 |
| 133594725 | 2011.09.15 17:30 | sell | 1.09 | gbpusd | 1.58332 | 1.58331 | 1.57902 | 2011.09.15 17:31 | 1.58331 | 0.00 | 0.00 | 0.00 | 1.09 |
| 132641432 | 2011.09.12 00:01 | sell | 1.02 | gbpusd | 1.58331 | 1.67415 | 1.57965 | 2011.09.12 08:11 | 1.58255 | 0.00 | 0.00 | 0.00 | 77.52 |
| 132641488 | 2011.09.12 00:01 | sell | 0.37 | gbpusd | 1.58327 | 1.67411 | 1.57961 | 2011.09.12 08:11 | 1.58260 | 0.00 | 0.00 | 0.00 | 24.79 |
| 133595140 | 2011.09.15 17:31 | sell | 0.66 | gbpusd | 1.58324 | 1.58244 | 1.57895 | 2011.09.15 17:37 | 1.58244 | 0.00 | 0.00 | 0.00 | 52.80 |
| 132798410 | 2011.09.12 16:46 | sell | 0.71 | gbpusd | 1.58308 | 1.58287 | 1.57878 | 2011.09.12 16:50 | 1.58287 | 0.00 | 0.00 | 0.00 | 14.91 |
| 132689938 | 2011.09.12 08:44 | sell | 0.23 | gbpusd | 1.58294 | 1.58264 | 1.57859 | 2011.09.12 09:16 | 1.58196 | 0.00 | 0.00 | 0.00 | 22.54 |
| 132686635 | 2011.09.12 08:18 | sell | 1.33 | gbpusd | 1.58294 | 1.67313 | 1.57863 | 2011.09.12 08:19 | 1.58279 | 0.00 | 0.00 | 0.00 | 19.95 |
| 132686620 | 2011.09.12 08:18 | sell | 2.24 | gbpusd | 1.58269 | 1.67291 | 1.57841 | 2011.09.12 08:19 | 1.58260 | 0.00 | 0.00 | 0.00 | 20.16 |
| 132814905 | 2011.09.12 17:44 | sell | 0.89 | gbpusd | 1.58262 | 1.58141 | 1.57830 | 2011.09.12 17:49 | 1.58103 | 0.00 | 0.00 | 0.00 | 141.51 |
| 132686815 | 2011.09.12 08:19 | buy | 2.23 | gbpusd | 1.58259 | 1.49238 | 1.58688 | 2011.09.12 08:36 | 1.58277 | 0.00 | 0.00 | 0.00 | 40.14 |
| 132685201 | 2011.09.12 08:11 | buy | 1.53 | gbpusd | 1.58259 | 1.58270 | 1.58690 | 2011.09.12 08:18 | 1.58270 | 0.00 | 0.00 | 0.00 | 16.83 |
| 132707341 | 2011.09.12 10:08 | sell | 0.26 | gbpusd | 1.58242 | 1.58225 | 1.57811 | 2011.09.12 10:12 | 1.58142 | 0.00 | 0.00 | 0.00 | 26.00 |
| 132686844 | 2011.09.12 08:20 | buy | 1.33 | gbpusd | 1.58242 | 1.58267 | 1.58672 | 2011.09.12 08:43 | 1.58289 | 0.00 | 0.00 | 0.00 | 62.51 |
| 132709500 | 2011.09.12 10:17 | sell | 0.26 | gbpusd | 1.58230 | 1.58192 | 1.57799 | 2011.09.12 10:28 | 1.58135 | 0.00 | 0.00 | 0.00 | 24.70 |
| 132695821 | 2011.09.12 09:16 | buy | 0.22 | gbpusd | 1.58199 | 1.58248 | 1.58629 | 2011.09.12 10:08 | 1.58248 | 0.00 | 0.00 | 0.00 | 10.78 |
| 132708446 | 2011.09.12 10:12 | buy | 0.26 | gbpusd | 1.58149 | 1.58234 | 1.58579 | 2011.09.12 10:17 | 1.58234 | 0.00 | 0.00 | 0.00 | 22.10 |
| 132712032 | 2011.09.12 10:28 | sell | 0.27 | gbpusd | 1.58123 | 1.58031 | 1.57691 | 2011.09.12 18:03 | 1.58021 | 0.00 | 0.00 | 0.00 | 27.54 |
| 132816717 | 2011.09.12 17:49 | sell | 1.12 | gbpusd | 1.58089 | 1.58045 | 1.57656 | 2011.09.12 18:01 | 1.57997 | 0.00 | 0.00 | 0.00 | 103.04 |
| 133606109 | 2011.09.15 18:11 | sell | 0.10 | gbpusd | 1.58065 | 1.58042 | 1.57646 | 2011.09.15 18:18 | 1.58042 | 0.00 | 0.00 | 0.00 | 2.30 |
| 133751605 | 2011.09.16 12:45 | sell | 0.34 | gbpusd | 1.58039 | 1.58013 | 1.57609 | 2011.09.16 14:39 | 1.57951 | 0.00 | 0.00 | 0.00 | 29.92 |
| 133607883 | 2011.09.15 18:18 | sell | 0.10 | gbpusd | 1.58010 | 1.57921 | 1.57591 | 2011.09.16 09:42 | 1.57921 | 0.00 | 0.00 | -0.33 | 8.90 |
| 132820878 | 2011.09.12 18:03 | sell | 1.32 | gbpusd | 1.58002 | 1.57923 | 1.57571 | 2011.09.12 18:07 | 1.57923 | 0.00 | 0.00 | 0.00 | 104.28 |
| 133776710 | 2011.09.16 14:39 | buy | 0.35 | gbpusd | 1.57945 | 1.48925 | 1.58375 | 2011.09.16 17:15 | 1.58375 | 0.00 | 0.00 | 0.00 | 150.50 |
| 133710671 | 2011.09.16 09:42 | buy | 0.26 | gbpusd | 1.57926 | 1.57964 | 1.58359 | 2011.09.16 12:44 | 1.58025 | 0.00 | 0.00 | 0.00 | 25.74 |
| 132994115 | 2011.09.13 11:26 | buy | 0.41 | gbpusd | 1.57907 | 1.57917 | 1.58325 | 2011.09.13 11:28 | 1.57917 | 0.00 | 0.00 | 0.00 | 4.10 |
| 132821933 | 2011.09.12 18:07 | sell | 1.94 | gbpusd | 1.57907 | 1.57816 | 1.57475 | 2011.09.12 18:09 | 1.57816 | 0.00 | 0.00 | 0.00 | 176.54 |
| 132994105 | 2011.09.13 11:26 | sell | 0.41 | gbpusd | 1.57875 | 1.57868 | 1.57457 | 2011.09.13 11:30 | 1.57747 | 0.00 | 0.00 | 0.00 | 52.48 |
| 132822741 | 2011.09.12 18:10 | sell | 0.78 | gbpusd | 1.57841 | 1.66860 | 1.57410 | 2011.09.13 11:08 | 1.57808 | 0.00 | 0.00 | -2.65 | 25.74 |
| 132822360 | 2011.09.12 18:09 | sell | 0.56 | gbpusd | 1.57829 | 1.57828 | 1.57397 | 2011.09.12 18:09 | 1.57828 | 0.00 | 0.00 | 0.00 | 0.56 |
| 132822525 | 2011.09.12 18:09 | sell | 1.35 | gbpusd | 1.57820 | 1.66841 | 1.57391 | 2011.09.13 11:08 | 1.57783 | 0.00 | 0.00 | -4.59 | 49.95 |
| 132988635 | 2011.09.13 11:08 | buy | 0.42 | gbpusd | 1.57776 | 1.57880 | 1.58194 | 2011.09.13 11:26 | 1.57880 | 0.00 | 0.00 | 0.00 | 43.68 |
| 132995071 | 2011.09.13 11:30 | buy | 0.41 | gbpusd | 1.57772 | 1.57781 | 1.58160 | 2011.09.13 11:30 | 1.57781 | 0.00 | 0.00 | 0.00 | 3.69 |
| 132995259 | 2011.09.13 11:30 | sell | 0.41 | gbpusd | 1.57764 | 1.66796 | 1.57346 | 2011.09.14 09:20 | 1.57346 | 0.00 | 0.00 | -1.35 | 171.38 |
| 134236469 | 2011.09.20 13:01 | sell | 0.30 | gbpusd | 1.57278 | 1.66295 | 1.56845 | 2011.09.20 17:08 | 1.56845 | 0.00 | 0.00 | 0.00 | 129.90 |
| 134234416 | 2011.09.20 12:53 | buy | 0.59 | gbpusd | 1.57259 | 1.57278 | 1.57691 | 2011.09.20 13:01 | 1.57278 | 0.00 | 0.00 | 0.00 | 11.21 |
| 134101854 | 2011.09.19 22:30 | sell | 0.37 | gbpusd | 1.57124 | 1.57090 | 1.56694 | 2011.09.19 22:51 | 1.57090 | 0.00 | 0.00 | 0.00 | 12.58 |
| 134106840 | 2011.09.19 22:51 | sell | 0.38 | gbpusd | 1.57069 | 1.57041 | 1.56639 | 2011.09.19 23:04 | 1.57041 | 0.00 | 0.00 | 0.00 | 10.64 |
| 134230475 | 2011.09.20 12:40 | buy | 0.29 | gbpusd | 1.57048 | 1.57185 | 1.57481 | 2011.09.20 12:46 | 1.57185 | 0.00 | 0.00 | 0.00 | 39.73 |
| 134090847 | 2011.09.19 21:24 | buy | 0.37 | gbpusd | 1.57039 | 1.57058 | 1.57471 | 2011.09.19 22:30 | 1.57112 | 0.00 | 0.00 | 0.00 | 27.01 |
| 134090844 | 2011.09.19 21:24 | sell | 0.37 | gbpusd | 1.57021 | 1.57016 | 1.56589 | 2011.09.19 21:44 | 1.57016 | 0.00 | 0.00 | 0.00 | 1.85 |
| 134108733 | 2011.09.19 23:04 | sell | 0.38 | gbpusd | 1.57003 | 1.56998 | 1.56580 | 2011.09.19 23:20 | 1.56998 | 0.00 | 0.00 | 0.00 | 1.90 |
| 134110225 | 2011.09.19 23:20 | sell | 0.38 | gbpusd | 1.56970 | 1.56964 | 1.56554 | 2011.09.20 01:07 | 1.56964 | 0.00 | 0.00 | -1.18 | 2.28 |
| 134022255 | 2011.09.19 16:18 | buy | 0.38 | gbpusd | 1.56948 | 1.57035 | 1.57379 | 2011.09.19 21:23 | 1.57035 | 0.00 | 0.00 | 0.00 | 33.06 |
| 134121981 | 2011.09.20 01:07 | sell | 0.38 | gbpusd | 1.56942 | 1.56742 | 1.56517 | 2011.09.20 01:24 | 1.56742 | 0.00 | 0.00 | 0.00 | 76.00 |
| 134022250 | 2011.09.19 16:18 | sell | 0.38 | gbpusd | 1.56929 | 1.56907 | 1.56498 | 2011.09.19 16:20 | 1.56848 | 0.00 | 0.00 | 0.00 | 30.78 |
| 134019094 | 2011.09.19 16:04 | buy | 0.38 | gbpusd | 1.56927 | 1.56937 | 1.57362 | 2011.09.19 16:18 | 1.56937 | 0.00 | 0.00 | 0.00 | 3.80 |
| 134127390 | 2011.09.20 01:26 | sell | 1.45 | gbpusd | 1.56751 | 1.56722 | 1.56333 | 2011.09.20 01:30 | 1.56722 | 0.00 | 0.00 | 0.00 | 42.05 |
| 134126311 | 2011.09.20 01:24 | sell | 3.28 | gbpusd | 1.56721 | 1.56643 | 1.56298 | 2011.09.20 01:41 | 1.56643 | 0.00 | 0.00 | 0.00 | 255.84 |
| 134129902 | 2011.09.20 01:42 | sell | 1.16 | gbpusd | 1.56622 | 1.56621 | 1.56204 | 2011.09.20 02:01 | 1.56621 | 0.00 | 0.00 | 0.00 | 1.16 |
| 134129834 | 2011.09.20 01:41 | sell | 3.00 | gbpusd | 1.56612 | 1.65647 | 1.56197 | 2011.09.20 02:04 | 1.56586 | 0.00 | 0.00 | 0.00 | 78.00 |
| 134132395 | 2011.09.20 02:04 | sell | 1.38 | gbpusd | 1.56569 | 1.65599 | 1.56149 | 2011.09.20 12:51 | 1.57210 | 0.00 | 0.00 | 0.00 | -884.58 |
| 134132372 | 2011.09.20 02:04 | sell | 2.33 | gbpusd | 1.56560 | 1.65583 | 1.56133 | 2011.09.20 12:39 | 1.56975 | 0.00 | 0.00 | 0.00 | -966.95 |
| 134032779 | 2011.09.19 16:49 | buy | 0.36 | gbpusd | 1.56508 | 1.56613 | 1.56939 | 2011.09.19 17:15 | 1.56613 | 0.00 | 0.00 | 0.00 | 37.80 |
| 135467619 | 2011.09.26 18:55 | buy | 0.73 | gbpusd | 1.55469 | 1.55472 | 1.55903 | 2011.09.26 19:00 | 1.55472 | 0.00 | 0.00 | 0.00 | 2.19 |
| 135468831 | 2011.09.26 19:00 | sell | 0.77 | gbpusd | 1.55463 | 1.64482 | 1.55032 | 2011.09.26 19:03 | 1.55422 | 0.00 | 0.00 | 0.00 | 31.57 |
| 135449459 | 2011.09.26 17:53 | sell | 0.91 | gbpusd | 1.55445 | 1.64464 | 1.55014 | 2011.09.26 17:55 | 1.55424 | 0.00 | 0.00 | 0.00 | 19.11 |
| 135451411 | 2011.09.26 17:59 | sell | 0.89 | gbpusd | 1.55428 | 1.64445 | 1.54995 | 2011.09.26 18:54 | 1.55406 | 0.00 | 0.00 | 0.00 | 19.58 |
| 135449386 | 2011.09.26 17:53 | sell | 1.51 | gbpusd | 1.55427 | 1.64447 | 1.54997 | 2011.09.26 17:56 | 1.55414 | 0.00 | 0.00 | 0.00 | 19.63 |
| 135451312 | 2011.09.26 17:58 | sell | 1.45 | gbpusd | 1.55403 | 1.64421 | 1.54971 | 2011.09.26 19:10 | 1.55380 | 0.00 | 0.00 | 0.00 | 33.35 |
| 135450542 | 2011.09.26 17:56 | buy | 1.42 | gbpusd | 1.55401 | 1.55419 | 1.55834 | 2011.09.26 17:58 | 1.55419 | 0.00 | 0.00 | 0.00 | 25.56 |
| 135447419 | 2011.09.26 17:47 | buy | 0.50 | gbpusd | 1.55289 | 1.55428 | 1.55720 | 2011.09.26 17:52 | 1.55428 | 0.00 | 0.00 | 0.00 | 69.50 |
| 135446294 | 2011.09.26 17:44 | buy | 1.86 | gbpusd | 1.55228 | 1.55431 | 1.55659 | 2011.09.26 17:52 | 1.55431 | 0.00 | 0.00 | 0.00 | 377.58 |
| 135443451 | 2011.09.26 17:34 | buy | 1.26 | gbpusd | 1.55162 | 1.55215 | 1.55592 | 2011.09.26 17:43 | 1.55215 | 0.00 | 0.00 | 0.00 | 66.78 |
| 135441137 | 2011.09.26 17:27 | buy | 0.30 | gbpusd | 1.55123 | 1.55220 | 1.55554 | 2011.09.26 17:43 | 1.55220 | 0.00 | 0.00 | 0.00 | 29.10 |
| 135440355 | 2011.09.26 17:24 | buy | 2.50 | gbpusd | 1.55107 | 1.55117 | 1.55538 | 2011.09.26 17:31 | 1.55117 | 0.00 | 0.00 | 0.00 | 25.00 |
| 129107656 | 2011.08.17 12:47 | sell | 0.14 | eurusd | 1.44343 | 1.44335 | 0.00000 | 2011.08.17 12:58 | 1.44335 | 0.00 | 0.00 | 0.00 | 1.12 |
| 129041334 | 2011.08.17 08:19 | sell | 0.22 | eurusd | 1.44168 | 1.44155 | 1.43730 | 2011.08.17 08:23 | 1.44155 | 0.00 | 0.00 | 0.00 | 2.86 |
| 129042048 | 2011.08.17 08:23 | sell | 0.22 | eurusd | 1.44154 | 1.44124 | 1.43716 | 2011.08.17 08:24 | 1.44085 | 0.00 | 0.00 | 0.00 | 15.18 |
| 129619831 | 2011.08.19 19:35 | sell | 0.59 | eurusd | 1.44098 | 1.44095 | 1.43660 | 2011.08.19 19:53 | 1.44054 | 0.00 | 0.00 | 0.00 | 25.96 |
| 129042303 | 2011.08.17 08:24 | sell | 0.23 | eurusd | 1.44074 | 1.44062 | 1.43637 | 2011.08.17 08:53 | 1.44062 | 0.00 | 0.00 | 0.00 | 2.76 |
| 128982837 | 2011.08.16 21:27 | buy | 0.23 | eurusd | 1.44064 | 1.44075 | 1.44501 | 2011.08.16 21:50 | 1.44075 | 0.00 | 0.00 | 0.00 | 2.53 |
| 129614887 | 2011.08.19 18:51 | sell | 0.27 | eurusd | 1.44036 | 1.44022 | 1.43597 | 2011.08.19 20:29 | 1.43951 | 0.00 | 0.00 | 0.00 | 22.95 |
| 128931919 | 2011.08.16 17:54 | sell | 0.13 | eurusd | 1.43976 | 1.43956 | 1.43537 | 2011.08.16 18:20 | 1.43956 | 0.00 | 0.00 | 0.00 | 2.60 |
| 129636519 | 2011.08.19 22:46 | sell | 0.11 | eurusd | 1.43970 | 1.43962 | 1.43534 | 2011.08.19 23:36 | 1.43962 | 0.00 | 0.00 | 0.00 | 0.88 |
| 129624757 | 2011.08.19 20:29 | buy | 0.13 | eurusd | 1.43958 | 1.43973 | 1.44396 | 2011.08.19 22:46 | 1.43973 | 0.00 | 0.00 | 0.00 | 1.95 |
| 128938087 | 2011.08.16 18:20 | sell | 0.14 | eurusd | 1.43952 | 1.43949 | 1.43513 | 2011.08.16 18:22 | 1.43949 | 0.00 | 0.00 | 0.00 | 0.42 |
| 128938549 | 2011.08.16 18:22 | sell | 0.14 | eurusd | 1.43940 | 1.43930 | 1.43501 | 2011.08.16 19:31 | 1.43930 | 0.00 | 0.00 | 0.00 | 1.40 |
| 128958763 | 2011.08.16 19:31 | sell | 0.24 | eurusd | 1.43921 | 1.43895 | 1.43486 | 2011.08.16 19:33 | 1.43895 | 0.00 | 0.00 | 0.00 | 6.24 |
| 128959338 | 2011.08.16 19:33 | sell | 0.24 | eurusd | 1.43879 | 1.43878 | 1.43443 | 2011.08.16 20:11 | 1.43878 | 0.00 | 0.00 | 0.00 | 0.24 |
| 128969682 | 2011.08.16 20:11 | sell | 0.24 | eurusd | 1.43842 | 1.43829 | 1.43406 | 2011.08.16 20:22 | 1.43829 | 0.00 | 0.00 | 0.00 | 3.12 |
| 128971842 | 2011.08.16 20:23 | sell | 0.24 | eurusd | 1.43818 | 1.43778 | 1.43379 | 2011.08.17 10:00 | 1.43778 | 0.00 | 0.00 | -1.32 | 9.60 |
| 129059702 | 2011.08.17 10:00 | sell | 0.24 | eurusd | 1.43767 | 1.43749 | 1.43330 | 2011.08.17 10:01 | 1.43749 | 0.00 | 0.00 | 0.00 | 4.32 |
| 129076307 | 2011.08.17 10:49 | sell | 0.22 | eurusd | 1.43762 | 1.52773 | 1.43323 | 2011.08.18 16:37 | 1.43323 | 0.00 | 0.00 | -3.63 | 96.58 |
| 129060598 | 2011.08.17 10:01 | sell | 0.24 | eurusd | 1.43736 | 1.43696 | 1.43299 | 2011.08.17 10:04 | 1.43696 | 0.00 | 0.00 | 0.00 | 9.60 |
| 129062098 | 2011.08.17 10:04 | sell | 0.25 | eurusd | 1.43677 | 1.43394 | 1.43240 | 2011.08.17 10:07 | 1.43394 | 0.00 | 0.00 | 0.00 | 70.75 |
| 129064091 | 2011.08.17 10:07 | sell | 0.25 | eurusd | 1.43386 | 1.43321 | 1.42946 | 2011.08.17 10:13 | 1.43321 | 0.00 | 0.00 | 0.00 | 16.25 |
| 129066572 | 2011.08.17 10:13 | sell | 0.25 | eurusd | 1.43312 | 1.52326 | 1.42876 | 2011.08.18 17:15 | 1.42876 | 0.00 | 0.00 | -4.13 | 109.00 |
| 128365343 | 2011.08.11 17:59 | buy | 0.30 | eurusd | 1.42691 | 1.42851 | 1.43129 | 2011.08.11 18:01 | 1.42851 | 0.00 | 0.00 | 0.00 | 48.00 |
| 128365334 | 2011.08.11 17:59 | sell | 0.30 | eurusd | 1.42679 | 1.51691 | 1.42241 | 2011.08.11 18:19 | 1.42513 | 0.00 | 0.00 | 0.00 | 49.80 |
| 128365159 | 2011.08.11 17:59 | buy | 0.35 | eurusd | 1.42632 | 1.42633 | 1.43069 | 2011.08.11 17:59 | 1.42680 | 0.00 | 0.00 | 0.00 | 16.80 |
| 128363878 | 2011.08.11 17:52 | sell | 0.35 | eurusd | 1.42565 | 1.51579 | 1.42129 | 2011.08.11 18:19 | 1.42511 | 0.00 | 0.00 | 0.00 | 18.90 |
| 128362843 | 2011.08.11 17:49 | buy | 0.27 | eurusd | 1.42555 | 1.42576 | 1.42993 | 2011.08.11 17:52 | 1.42576 | 0.00 | 0.00 | 0.00 | 5.67 |
| 128371114 | 2011.08.11 18:19 | buy | 0.37 | eurusd | 1.42518 | 1.33505 | 1.42955 | 2011.08.15 03:22 | 1.42955 | 0.00 | 0.00 | -0.15 | 161.69 |
| 128371402 | 2011.08.11 18:21 | buy | 0.27 | eurusd | 1.42486 | 1.33473 | 1.42923 | 2011.08.15 03:03 | 1.42923 | 0.00 | 0.00 | -0.11 | 117.99 |
| 128263543 | 2011.08.11 10:51 | sell | 0.12 | eurusd | 1.42418 | 1.42338 | 1.41979 | 2011.08.11 12:19 | 1.42338 | 0.00 | 0.00 | 0.00 | 9.60 |
| 128262703 | 2011.08.11 10:47 | buy | 0.13 | eurusd | 1.42414 | 1.42471 | 1.42852 | 2011.08.11 10:50 | 1.42471 | 0.00 | 0.00 | 0.00 | 7.41 |
| 128259807 | 2011.08.11 10:36 | buy | 0.14 | eurusd | 1.42377 | 1.42493 | 1.42814 | 2011.08.11 10:41 | 1.42493 | 0.00 | 0.00 | 0.00 | 16.24 |
| 128126560 | 2011.08.10 16:49 | buy | 0.14 | eurusd | 1.42372 | 1.42380 | 1.42810 | 2011.08.11 10:36 | 1.42380 | 0.00 | 0.00 | 0.13 | 1.12 |
| 128255489 | 2011.08.11 10:24 | sell | 0.27 | eurusd | 1.42367 | 1.42360 | 1.41931 | 2011.08.11 10:25 | 1.42309 | 0.00 | 0.00 | 0.00 | 15.66 |
| 128280850 | 2011.08.11 12:25 | sell | 0.10 | eurusd | 1.42365 | 1.42305 | 1.41926 | 2011.08.11 12:46 | 1.42161 | 0.00 | 0.00 | 0.00 | 20.40 |
| 128259801 | 2011.08.11 10:36 | sell | 0.14 | eurusd | 1.42364 | 1.42333 | 1.41927 | 2011.08.11 12:17 | 1.42333 | 0.00 | 0.00 | 0.00 | 4.34 |
| 128125812 | 2011.08.10 16:45 | buy | 0.13 | eurusd | 1.42336 | 1.42356 | 1.42774 | 2011.08.10 16:49 | 1.42356 | 0.00 | 0.00 | 0.00 | 2.60 |
| 128280812 | 2011.08.11 12:25 | sell | 0.15 | eurusd | 1.42335 | 1.42280 | 1.41898 | 2011.08.11 12:45 | 1.42174 | 0.00 | 0.00 | 0.00 | 24.15 |
| 128279288 | 2011.08.11 12:17 | buy | 0.10 | eurusd | 1.42332 | 1.42345 | 1.42769 | 2011.08.11 12:25 | 1.42345 | 0.00 | 0.00 | 0.00 | 1.30 |
| 128175859 | 2011.08.10 21:39 | buy | 0.18 | eurusd | 1.42315 | 1.42367 | 1.42751 | 2011.08.11 10:23 | 1.42407 | 0.00 | 0.00 | 0.16 | 16.56 |
| 128174925 | 2011.08.10 21:29 | buy | 0.24 | eurusd | 1.42296 | 1.42314 | 1.42732 | 2011.08.10 21:36 | 1.42314 | 0.00 | 0.00 | 0.00 | 4.32 |
| 128256281 | 2011.08.11 10:25 | buy | 0.14 | eurusd | 1.42294 | 1.42380 | 1.42731 | 2011.08.11 10:36 | 1.42380 | 0.00 | 0.00 | 0.00 | 12.04 |
| 128171400 | 2011.08.10 21:11 | buy | 0.32 | eurusd | 1.42264 | 1.42306 | 1.42700 | 2011.08.10 21:14 | 1.42343 | 0.00 | 0.00 | 0.00 | 25.28 |
| 128172953 | 2011.08.10 21:17 | buy | 0.16 | eurusd | 1.42257 | 1.42307 | 1.42693 | 2011.08.10 21:24 | 1.42307 | 0.00 | 0.00 | 0.00 | 8.00 |
| 128284235 | 2011.08.11 12:45 | buy | 0.10 | eurusd | 1.42167 | 1.42498 | 1.42604 | 2011.08.11 17:50 | 1.42498 | 0.00 | 0.00 | 0.00 | 33.10 |
| 128300830 | 2011.08.11 13:59 | buy | 0.10 | eurusd | 1.42094 | 1.42473 | 1.42531 | 2011.08.11 17:48 | 1.42531 | 0.00 | 0.00 | 0.00 | 43.70 |
| 128300477 | 2011.08.11 13:57 | sell | 0.10 | eurusd | 1.42089 | 1.42087 | 1.41651 | 2011.08.11 13:59 | 1.42087 | 0.00 | 0.00 | 0.00 | 0.20 |
| 128285713 | 2011.08.11 12:51 | buy | 0.10 | eurusd | 1.42082 | 1.42094 | 1.42520 | 2011.08.11 13:57 | 1.42094 | 0.00 | 0.00 | 0.00 | 1.20 |
| 128167959 | 2011.08.10 20:50 | buy | 0.65 | eurusd | 1.42042 | 1.42045 | 1.42481 | 2011.08.10 21:00 | 1.42045 | 0.00 | 0.00 | 0.00 | 1.95 |
| 128146459 | 2011.08.10 18:23 | sell | 0.12 | euraud | 1.39025 | 1.48080 | 1.38630 | 2011.08.10 19:59 | 1.38630 | 0.00 | 0.00 | 0.00 | 48.55 |
| 128162717 | 2011.08.10 20:16 | sell | 0.12 | euraud | 1.38622 | 1.38555 | 1.38226 | 2011.08.10 20:30 | 1.38555 | 0.00 | 0.00 | 0.00 | 8.22 |
| 128131194 | 2011.08.10 17:01 | sell | 0.13 | euraud | 1.38605 | 1.38516 | 1.38206 | 2011.08.10 17:16 | 1.38455 | 0.00 | 0.00 | 0.00 | 19.97 |
| 128125793 | 2011.08.10 16:45 | buy | 0.14 | euraud | 1.38598 | 1.38612 | 1.39003 | 2011.08.10 17:01 | 1.38612 | 0.00 | 0.00 | 0.00 | 2.00 |
| 128165766 | 2011.08.10 20:38 | sell | 0.12 | euraud | 1.38578 | 1.38506 | 1.38182 | 2011.08.10 20:49 | 1.38506 | 0.00 | 0.00 | 0.00 | 8.86 |
| 128168933 | 2011.08.10 20:59 | sell | 0.15 | euraud | 1.38507 | 1.47548 | 1.38098 | 2011.08.10 21:11 | 1.38341 | 0.00 | 0.00 | 0.00 | 25.62 |
| 128132904 | 2011.08.10 17:23 | sell | 0.13 | euraud | 1.38504 | 1.38469 | 1.38104 | 2011.08.10 20:52 | 1.38469 | 0.00 | 0.00 | 0.00 | 4.66 |
| 128131779 | 2011.08.10 17:14 | buy | 0.12 | euraud | 1.38496 | 1.38508 | 1.38891 | 2011.08.10 17:23 | 1.38508 | 0.00 | 0.00 | 0.00 | 1.48 |
| 128167954 | 2011.08.10 20:50 | sell | 1.02 | euraud | 1.38471 | 1.38470 | 1.38072 | 2011.08.10 20:52 | 1.38436 | 0.00 | 0.00 | 0.00 | 36.65 |
| 135441440 | 2011.09.26 17:28 | sell | 0.22 | euraud | 1.38460 | 1.38247 | 1.38052 | 2011.09.26 20:56 | 1.38247 | 0.00 | 0.00 | 0.00 | 45.61 |
| 128168253 | 2011.08.10 20:52 | sell | 0.75 | euraud | 1.38432 | 1.47478 | 1.38028 | 2011.08.10 21:11 | 1.38351 | 0.00 | 0.00 | 0.00 | 62.47 |
| 128171387 | 2011.08.10 21:11 | sell | 0.52 | euraud | 1.38264 | 1.47364 | 1.37914 | 2011.08.11 22:05 | 1.37914 | 0.00 | 0.00 | 8.91 | 187.75 |
| 128171367 | 2011.08.10 21:11 | sell | 0.85 | euraud | 1.38253 | 1.47353 | 1.37903 | 2011.08.11 22:05 | 1.37903 | 0.00 | 0.00 | 14.57 | 306.91 |
| 129614497 | 2011.08.19 18:49 | sell | 0.28 | euraud | 1.37998 | 1.47078 | 1.37628 | 2011.08.23 06:58 | 1.37628 | 0.00 | 0.00 | 3.81 | 108.27 |
| 129619555 | 2011.08.19 19:32 | sell | 1.55 | euraud | 1.37984 | 1.47064 | 1.37614 | 2011.08.23 07:00 | 1.37614 | 0.00 | 0.00 | 21.12 | 599.36 |
| 133734089 | 2011.09.16 11:11 | buy | 0.26 | eurusd | 1.37807 | 1.37879 | 1.38245 | 2011.09.16 11:16 | 1.37879 | 0.00 | 0.00 | 0.00 | 18.72 |
| 128419441 | 2011.08.11 22:57 | sell | 1.57 | euraud | 1.37577 | 1.46657 | 1.37207 | 2011.08.15 03:07 | 1.37207 | 0.00 | 0.00 | 20.31 | 605.27 |
| 128419437 | 2011.08.11 22:57 | sell | 1.57 | euraud | 1.37568 | 1.46648 | 1.37198 | 2011.08.15 03:07 | 1.37198 | 0.00 | 0.00 | 20.31 | 605.28 |
| 128947719 | 2011.08.16 19:09 | sell | 0.12 | euraud | 1.37524 | 1.46575 | 1.37125 | 2011.08.17 08:24 | 1.37403 | 0.00 | 0.00 | 0.74 | 15.24 |
| 128931536 | 2011.08.16 17:52 | sell | 0.14 | euraud | 1.37370 | 1.37310 | 1.36962 | 2011.08.16 18:00 | 1.37310 | 0.00 | 0.00 | 0.00 | 8.81 |
| 129045417 | 2011.08.17 08:53 | buy | 0.23 | euraud | 1.37362 | 1.28282 | 1.37732 | 2011.08.18 11:06 | 1.37732 | 0.00 | 0.00 | -17.68 | 89.15 |
| 128933633 | 2011.08.16 18:00 | sell | 0.13 | euraud | 1.37275 | 1.46316 | 1.36866 | 2011.08.17 10:12 | 1.37105 | 0.00 | 0.00 | 0.81 | 23.13 |
| 129076313 | 2011.08.17 10:49 | buy | 0.21 | euraud | 1.37150 | 1.28070 | 1.37520 | 2011.08.17 11:44 | 1.37250 | 0.00 | 0.00 | 0.00 | 22.02 |
| 134233000 | 2011.09.20 12:46 | buy | 0.13 | eurusd | 1.37143 | 1.28125 | 1.37575 | 2011.09.20 12:52 | 1.37052 | 0.00 | 0.00 | 0.00 | -11.83 |
| 134234387 | 2011.09.20 12:53 | buy | 2.19 | eurusd | 1.37128 | 1.37129 | 1.37558 | 2011.09.20 13:02 | 1.37129 | 0.00 | 0.00 | 0.00 | 2.19 |
| 134235192 | 2011.09.20 12:56 | buy | 0.14 | eurusd | 1.37097 | 1.37114 | 1.37527 | 2011.09.20 13:03 | 1.37114 | 0.00 | 0.00 | 0.00 | 2.38 |
| 132777793 | 2011.09.12 15:26 | sell | 1.22 | eurusd | 1.36711 | 1.36587 | 1.36275 | 2011.09.12 15:34 | 1.36571 | 0.00 | 0.00 | 0.00 | 170.80 |
| 132774636 | 2011.09.12 15:14 | buy | 1.37 | eurusd | 1.36606 | 1.36686 | 1.37044 | 2011.09.12 15:26 | 1.36724 | 0.00 | 0.00 | 0.00 | 161.66 |
| 132779909 | 2011.09.12 15:34 | sell | 1.07 | eurusd | 1.36598 | 1.36596 | 1.36159 | 2011.09.12 16:12 | 1.36560 | 0.00 | 0.00 | 0.00 | 40.66 |
| 132774625 | 2011.09.12 15:14 | sell | 1.37 | eurusd | 1.36594 | 1.45606 | 1.36156 | 2011.09.12 15:15 | 1.36538 | 0.00 | 0.00 | 0.00 | 76.72 |

| Ticket | Open Time | Type | Size | Symbol | Price | S/L | T/P | Close Time | Price | Commission | Taxes | Swap | Profit |
|---|---|---|---|---|---|---|---|---|---|---|---|---|---|
| 132790707 | 2011.09.12 16:18 | sell | 0.73 | eurusd | 1.36589 | 1.36473 | 1.36151 | 2011.09.12 16:25 | 1.36473 | 0.00 | 0.00 | 0.00 | 84.68 |
| 132779844 | 2011.09.12 15:34 | sell | 1.67 | eurusd | 1.36567 | 1.36563 | 1.36128 | 2011.09.12 16:14 | 1.36563 | 0.00 | 0.00 | 0.00 | 6.68 |
| 132789719 | 2011.09.12 16:14 | sell | 1.93 | eurusd | 1.36554 | 1.36463 | 1.36115 | 2011.09.12 16:24 | 1.36463 | 0.00 | 0.00 | 0.00 | 175.63 |
| 132774982 | 2011.09.12 15:15 | buy | 0.90 | eurusd | 1.36534 | 1.27523 | 1.36973 | 2011.09.12 15:26 | 1.36703 | 0.00 | 0.00 | 0.00 | 152.10 |
| 132767193 | 2011.09.12 14:49 | buy | 2.02 | eurusd | 1.36522 | 1.36590 | 1.36961 | 2011.09.12 15:14 | 1.36590 | 0.00 | 0.00 | 0.00 | 137.36 |
| 132767187 | 2011.09.12 14:49 | sell | 2.02 | eurusd | 1.36511 | 1.45522 | 1.36072 | 2011.09.12 14:49 | 1.36496 | 0.00 | 0.00 | 0.00 | 30.30 |
| 132792520 | 2011.09.12 16:25 | sell | 1.53 | eurusd | 1.36469 | 1.36389 | 1.36030 | 2011.09.12 16:33 | 1.36389 | 0.00 | 0.00 | 0.00 | 122.40 |
| 132767784 | 2011.09.12 14:52 | buy | 0.82 | eurusd | 1.36449 | 1.36587 | 1.36887 | 2011.09.12 15:14 | 1.36587 | 0.00 | 0.00 | 0.00 | 113.16 |
| 133021766 | 2011.09.13 13:07 | sell | 0.34 | eurusd | 1.36441 | 1.36400 | 1.36004 | 2011.09.13 13:10 | 1.36400 | 0.00 | 0.00 | 0.00 | 13.94 |
| 133023359 | 2011.09.13 13:11 | sell | 0.34 | eurusd | 1.36436 | 1.36349 | 1.35996 | 2011.09.13 13:19 | 1.36349 | 0.00 | 0.00 | 0.00 | 29.58 |
| 133025243 | 2011.09.13 13:19 | sell | 0.34 | eurusd | 1.36435 | 1.36419 | 1.35996 | 2011.09.13 14:08 | 1.36419 | 0.00 | 0.00 | 0.00 | 5.44 |
| 132742513 | 2011.09.12 12:49 | sell | 0.24 | eurusd | 1.36388 | 1.36365 | 1.35950 | 2011.09.12 12:56 | 1.36365 | 0.00 | 0.00 | 0.00 | 5.52 |
| 132794657 | 2011.09.12 16:33 | sell | 2.07 | eurusd | 1.36379 | 1.36292 | 1.35939 | 2011.09.12 16:36 | 1.36292 | 0.00 | 0.00 | 0.00 | 180.09 |
| 132744296 | 2011.09.12 12:57 | buy | 0.25 | eurusd | 1.36367 | 1.36516 | 1.36807 | 2011.09.12 14:49 | 1.36516 | 0.00 | 0.00 | 0.00 | 37.25 |
| 134127446 | 2011.09.20 01:26 | sell | 0.78 | eurusd | 1.36283 | 1.36171 | 1.35855 | 2011.09.20 01:27 | 1.36171 | 0.00 | 0.00 | 0.00 | 87.36 |
| 132795609 | 2011.09.12 16:36 | sell | 2.16 | eurusd | 1.36275 | 1.36231 | 1.35836 | 2011.09.12 16:39 | 1.36231 | 0.00 | 0.00 | 0.00 | 95.04 |
| 134126356 | 2011.09.20 01:24 | sell | 1.44 | eurusd | 1.36256 | 1.36140 | 1.35828 | 2011.09.20 01:25 | 1.36140 | 0.00 | 0.00 | 0.00 | 167.04 |
| 132796560 | 2011.09.12 16:39 | sell | 1.46 | eurusd | 1.36251 | 1.45263 | 1.35813 | 2011.09.12 17:50 | 1.36195 | 0.00 | 0.00 | 0.00 | 81.76 |
| 132818333 | 2011.09.12 17:56 | sell | 1.04 | eurusd | 1.36234 | 1.36129 | 1.35794 | 2011.09.12 18:00 | 1.36129 | 0.00 | 0.00 | 0.00 | 109.20 |
| 132796444 | 2011.09.12 16:39 | sell | 2.26 | eurusd | 1.36223 | 1.36124 | 1.35783 | 2011.09.12 17:59 | 1.36088 | 0.00 | 0.00 | 0.00 | 305.10 |
| 132884093 | 2011.09.12 22:45 | sell | 0.94 | eurusd | 1.36217 | 1.45230 | 1.35780 | 2011.09.13 10:12 | 1.36171 | 0.00 | 0.00 | -4.98 | 43.24 |
| 132973554 | 2011.09.13 10:18 | sell | 0.29 | eurusd | 1.36195 | 1.36171 | 1.35756 | 2011.09.13 10:38 | 1.36110 | 0.00 | 0.00 | 0.00 | 24.65 |
| 132869915 | 2011.09.12 21:43 | buy | 1.04 | eurusd | 1.36191 | 1.36204 | 1.36628 | 2011.09.12 21:44 | 1.36204 | 0.00 | 0.00 | 0.00 | 13.52 |
| 132871220 | 2011.09.12 21:46 | buy | 1.02 | eurusd | 1.36190 | 1.36195 | 1.36627 | 2011.09.12 22:45 | 1.36195 | 0.00 | 0.00 | 0.00 | 5.10 |
| 134127785 | 2011.09.20 01:27 | buy | 0.93 | eurusd | 1.36189 | 1.27162 | 1.36612 | 2011.09.20 01:27 | 1.36218 | 0.00 | 0.00 | 0.00 | 26.97 |
| 132884055 | 2011.09.12 22:45 | sell | 1.48 | eurusd | 1.36185 | 1.45199 | 1.35749 | 2011.09.13 10:12 | 1.36167 | 0.00 | 0.00 | -7.84 | 26.64 |
| 132871215 | 2011.09.12 21:46 | sell | 1.02 | eurusd | 1.36177 | 1.45190 | 1.35740 | 2011.09.12 21:46 | 1.36159 | 0.00 | 0.00 | 0.00 | 18.36 |
| 132820648 | 2011.09.12 18:02 | sell | 1.76 | eurusd | 1.36168 | 1.36163 | 1.35728 | 2011.09.12 18:06 | 1.36163 | 0.00 | 0.00 | 0.00 | 8.80 |
| 132980144 | 2011.09.13 10:45 | sell | 0.29 | eurusd | 1.36162 | 1.36135 | 1.35723 | 2011.09.13 10:48 | 1.36073 | 0.00 | 0.00 | 0.00 | 25.81 |
| 132822028 | 2011.09.12 18:08 | sell | 1.15 | eurusd | 1.36154 | 1.36113 | 1.35713 | 2011.09.12 19:34 | 1.36113 | 0.00 | 0.00 | 0.00 | 47.15 |
| 134129838 | 2011.09.20 01:41 | sell | 1.80 | eurusd | 1.36153 | 1.36110 | 1.35719 | 2011.09.20 01:59 | 1.36061 | 0.00 | 0.00 | 0.00 | 165.60 |
| 132867413 | 2011.09.12 21:37 | sell | 0.94 | eurusd | 1.36151 | 1.45169 | 1.35719 | 2011.09.12 21:37 | 1.36124 | 0.00 | 0.00 | 0.00 | 25.38 |
| 132840308 | 2011.09.12 19:35 | sell | 0.98 | eurusd | 1.36145 | 1.35993 | 1.35704 | 2011.09.12 19:43 | 1.35993 | 0.00 | 0.00 | 0.00 | 148.96 |
| 132970810 | 2011.09.13 10:12 | buy | 0.28 | eurusd | 1.36142 | 1.36206 | 1.36579 | 2011.09.13 10:18 | 1.36206 | 0.00 | 0.00 | 0.00 | 17.92 |
| 133003021 | 2011.09.13 11:59 | sell | 0.39 | eurusd | 1.36128 | 1.36030 | 1.35688 | 2011.09.13 12:18 | 1.36030 | 0.00 | 0.00 | 0.00 | 38.22 |
| 132820024 | 2011.09.12 18:00 | sell | 2.02 | eurusd | 1.36125 | 1.45137 | 1.35687 | 2011.09.12 19:33 | 1.36082 | 0.00 | 0.00 | 0.00 | 86.86 |
| 132867583 | 2011.09.12 21:37 | buy | 0.98 | eurusd | 1.36123 | 1.36164 | 1.36560 | 2011.09.12 21:38 | 1.36164 | 0.00 | 0.00 | 0.00 | 40.18 |
| 132978578 | 2011.09.13 10:38 | buy | 0.30 | eurusd | 1.36120 | 1.36193 | 1.36560 | 2011.09.13 10:45 | 1.36193 | 0.00 | 0.00 | 0.00 | 21.90 |
| 132840090 | 2011.09.12 19:34 | sell | 1.56 | eurusd | 1.36102 | 1.36023 | 1.35663 | 2011.09.12 19:42 | 1.35992 | 0.00 | 0.00 | 0.00 | 171.60 |
| 134045252 | 2011.09.19 17:33 | sell | 0.38 | eurusd | 1.36091 | 1.36063 | 1.35654 | 2011.09.20 02:05 | 1.36063 | 0.00 | 0.00 | -1.71 | 10.64 |
| 134037853 | 2011.09.19 17:05 | sell | 0.37 | eurusd | 1.36087 | 1.36071 | 1.35648 | 2011.09.19 17:16 | 1.36014 | 0.00 | 0.00 | 0.00 | 27.01 |
| 132881398 | 2011.09.12 22:32 | sell | 1.17 | eurusd | 1.36084 | 1.45105 | 1.35655 | 2011.09.12 22:33 | 1.36068 | 0.00 | 0.00 | 0.00 | 18.72 |
| 132980900 | 2011.09.13 10:48 | buy | 0.30 | eurusd | 1.36071 | 1.36114 | 1.36509 | 2011.09.13 11:59 | 1.36114 | 0.00 | 0.00 | 0.00 | 12.90 |
| 132881733 | 2011.09.12 22:33 | buy | 0.95 | eurusd | 1.36058 | 1.36190 | 1.36494 | 2011.09.12 22:45 | 1.36190 | 0.00 | 0.00 | 0.00 | 125.40 |
| 134132532 | 2011.09.20 02:05 | sell | 0.76 | eurusd | 1.36051 | 1.36041 | 1.35614 | 2011.09.20 05:46 | 1.36041 | 0.00 | 0.00 | 0.00 | 7.60 |
| 133012746 | 2011.09.13 12:31 | sell | 0.40 | eurusd | 1.36034 | 1.36032 | 1.35596 | 2011.09.19 17:00 | 1.36032 | 0.00 | 0.00 | -12.24 | 0.80 |
| 134155126 | 2011.09.20 06:14 | sell | 0.56 | eurusd | 1.36030 | 1.45043 | 1.35593 | 2011.09.20 12:51 | 1.37079 | 0.00 | 0.00 | 0.00 | -587.44 |
| 134151674 | 2011.09.20 05:46 | sell | 0.61 | eurusd | 1.36028 | 1.36008 | 1.35596 | 2011.09.20 06:14 | 1.36008 | 0.00 | 0.00 | 0.00 | 12.20 |
| 133009509 | 2011.09.13 12:18 | sell | 0.39 | eurusd | 1.36026 | 1.35956 | 1.35588 | 2011.09.13 12:25 | 1.35918 | 0.00 | 0.00 | 0.00 | 42.12 |
| 132866876 | 2011.09.12 21:36 | buy | 1.49 | eurusd | 1.36018 | 1.36170 | 1.36454 | 2011.09.12 21:37 | 1.36170 | 0.00 | 0.00 | 0.00 | 226.48 |
| 134040305 | 2011.09.19 17:16 | buy | 0.37 | eurusd | 1.36012 | 1.36105 | 1.36452 | 2011.09.19 17:33 | 1.36105 | 0.00 | 0.00 | 0.00 | 34.41 |
| 132866864 | 2011.09.12 21:36 | sell | 1.49 | eurusd | 1.36004 | 1.45018 | 1.35568 | 2011.09.12 21:51 | 1.35992 | 0.00 | 0.00 | 0.00 | 17.88 |
| 134132013 | 2011.09.20 02:01 | sell | 1.32 | eurusd | 1.36002 | 1.45016 | 1.35566 | 2011.09.20 12:40 | 1.36985 | 0.00 | 0.00 | 0.00 | -1 297.56 |
| 132842344 | 2011.09.12 19:43 | sell | 1.53 | eurusd | 1.35990 | 1.35911 | 1.35549 | 2011.09.12 19:51 | 1.35911 | 0.00 | 0.00 | 0.00 | 120.87 |
| 132872458 | 2011.09.12 21:51 | buy | 1.01 | eurusd | 1.35989 | 1.36096 | 1.36425 | 2011.09.12 22:32 | 1.36096 | 0.00 | 0.00 | 0.00 | 108.07 |
| 134036464 | 2011.09.19 17:01 | buy | 0.36 | eurusd | 1.35981 | 1.36077 | 1.36420 | 2011.09.19 17:05 | 1.36077 | 0.00 | 0.00 | 0.00 | 34.56 |
| 132844309 | 2011.09.12 19:52 | sell | 1.13 | eurusd | 1.35941 | 1.44951 | 1.35501 | 2011.09.12 20:20 | 1.35925 | 0.00 | 0.00 | 0.00 | 18.08 |
| 133011166 | 2011.09.13 12:25 | buy | 0.40 | eurusd | 1.35939 | 1.35940 | 1.36378 | 2011.09.13 12:31 | 1.36030 | 0.00 | 0.00 | 0.00 | 36.40 |
| 132849848 | 2011.09.12 20:20 | buy | 1.03 | eurusd | 1.35922 | 1.36004 | 1.36360 | 2011.09.12 21:36 | 1.36004 | 0.00 | 0.00 | 0.00 | 84.46 |
| 132844121 | 2011.09.12 19:51 | sell | 1.48 | eurusd | 1.35904 | 1.44915 | 1.35465 | 2011.09.12 20:22 | 1.35892 | 0.00 | 0.00 | 0.00 | 17.76 |
| 132641555 | 2011.09.12 00:01 | buy | 0.22 | eurusd | 1.35889 | 1.36168 | 1.36289 | 2011.09.12 12:45 | 1.36168 | 0.00 | 0.00 | 0.00 | 61.38 |
| 132641582 | 2011.09.12 00:01 | buy | 0.14 | eurusd | 1.35888 | 1.36166 | 1.36288 | 2011.09.12 12:48 | 1.36288 | 0.00 | 0.00 | 0.00 | 56.00 |
| 132850413 | 2011.09.12 20:22 | buy | 1.54 | eurusd | 1.35842 | 1.36003 | 1.36280 | 2011.09.12 21:36 | 1.36003 | 0.00 | 0.00 | 0.00 | 247.94 |
| 135458097 | 2011.09.26 18:17 | buy | 0.66 | eurusd | 1.35046 | 1.35060 | 1.35483 | 2011.09.26 18:20 | 1.35060 | 0.00 | 0.00 | 0.00 | 9.24 |
| 135458091 | 2011.09.26 18:17 | sell | 0.66 | eurusd | 1.35033 | 1.44046 | 1.34596 | 2011.09.26 19:03 | 1.34955 | 0.00 | 0.00 | 0.00 | 51.48 |
| 135457055 | 2011.09.26 18:14 | buy | 0.66 | eurusd | 1.35029 | 1.35033 | 1.35468 | 2011.09.26 18:17 | 1.35033 | 0.00 | 0.00 | 0.00 | 2.64 |
| 135457048 | 2011.09.26 18:14 | sell | 0.66 | eurusd | 1.35018 | 1.44029 | 1.34579 | 2011.09.26 18:14 | 1.34990 | 0.00 | 0.00 | 0.00 | 18.48 |
| 135455352 | 2011.09.26 18:09 | buy | 0.66 | eurusd | 1.35001 | 1.35002 | 1.35441 | 2011.09.26 18:10 | 1.35002 | 0.00 | 0.00 | 0.00 | 0.66 |
| 135454539 | 2011.09.26 18:07 | sell | 0.65 | eurusd | 1.34984 | 1.43996 | 1.34546 | 2011.09.26 18:09 | 1.34953 | 0.00 | 0.00 | 0.00 | 20.15 |
| 135451303 | 2011.09.26 17:58 | buy | 1.60 | eurusd | 1.34760 | 1.34932 | 1.35199 | 2011.09.26 18:07 | 1.34948 | 0.00 | 0.00 | 0.00 | 300.80 |
| 135450519 | 2011.09.26 17:56 | buy | 1.55 | eurusd | 1.34736 | 1.34767 | 1.35175 | 2011.09.26 17:58 | 1.34767 | 0.00 | 0.00 | 0.00 | 48.05 |
| 135447369 | 2011.09.26 17:47 | buy | 0.76 | eurusd | 1.34641 | 1.34763 | 1.35080 | 2011.09.26 17:54 | 1.34763 | 0.00 | 0.00 | 0.00 | 92.72 |
| 135447364 | 2011.09.26 17:47 | sell | 0.76 | eurusd | 1.34630 | 1.43641 | 1.34191 | 2011.09.26 17:48 | 1.34607 | 0.00 | 0.00 | 0.00 | 17.48 |
| 135446210 | 2011.09.26 17:43 | sell | 1.26 | eurusd | 1.34608 | 1.43617 | 1.34167 | 2011.09.26 17:43 | 1.34589 | 0.00 | 0.00 | 0.00 | 23.94 |
| 135447206 | 2011.09.26 17:46 | sell | 1.16 | eurusd | 1.34599 | 1.43611 | 1.34161 | 2011.09.26 20:01 | 1.34577 | 0.00 | 0.00 | 0.00 | 25.52 |
| 135446302 | 2011.09.26 17:44 | buy | 1.13 | eurusd | 1.34587 | 1.34602 | 1.35025 | 2011.09.26 17:46 | 1.34602 | 0.00 | 0.00 | 0.00 | 16.95 |
| 135446380 | 2011.09.26 17:44 | buy | 0.73 | eurusd | 1.34555 | 1.34599 | 1.34996 | 2011.09.26 17:46 | 1.34599 | 0.00 | 0.00 | 0.00 | 32.12 |
| 133595129 | 2011.09.15 17:31 | sell | 1.04 | euraud | 1.34496 | 1.43576 | 1.34126 | 2011.09.15 18:35 | 1.34319 | 0.00 | 0.00 | 0.00 | 189.92 |
| 135443129 | 2011.09.26 17:33 | sell | 1.93 | eurusd | 1.34491 | 1.43503 | 1.34053 | 2011.09.26 17:34 | 1.34471 | 0.00 | 0.00 | 0.00 | 38.60 |
| 135443255 | 2011.09.26 17:34 | buy | 1.94 | eurusd | 1.34480 | 1.34562 | 1.34920 | 2011.09.26 17:43 | 1.34606 | 0.00 | 0.00 | 0.00 | 244.44 |
| 135440961 | 2011.09.26 17:26 | buy | 0.64 | eurusd | 1.34444 | 1.34499 | 1.34883 | 2011.09.26 17:33 | 1.34499 | 0.00 | 0.00 | 0.00 | 35.20 |
| 135443501 | 2011.09.26 17:34 | buy | 0.76 | eurusd | 1.34442 | 1.34445 | 1.34882 | 2011.09.26 17:37 | 1.34445 | 0.00 | 0.00 | 0.00 | 2.28 |
| 135440362 | 2011.09.26 17:24 | buy | 1.52 | eurusd | 1.34414 | 1.34435 | 1.34854 | 2011.09.26 17:31 | 1.34435 | 0.00 | 0.00 | 0.00 | 31.92 |
| 134130306 | 2011.09.20 01:46 | buy | 0.37 | euraud | 1.34028 | 1.24928 | 1.34378 | 2011.09.20 12:51 | 1.33230 | 0.00 | 0.00 | 0.00 | -303.68 |
| 134127420 | 2011.09.20 01:26 | buy | 0.83 | euraud | 1.33899 | 1.24811 | 1.34261 | 2011.09.20 01:35 | 1.33958 | 0.00 | 0.00 | 0.00 | 49.78 |
| 134234401 | 2011.09.20 12:53 | buy | 1.44 | euraud | 1.33255 | 1.24209 | 1.33659 | 2011.09.21 02:10 | 1.33659 | 0.00 | 0.00 | -33.05 | 595.28 |
| 134234409 | 2011.09.20 12:53 | buy | 0.93 | euraud | 1.33254 | 1.24208 | 1.33658 | 2011.09.21 02:10 | 1.33658 | 0.00 | 0.00 | -21.34 | 384.47 |
| 134232989 | 2011.09.20 12:46 | buy | 0.23 | euraud | 1.33187 | 1.33206 | 1.33590 | 2011.09.20 12:52 | 1.33206 | 0.00 | 0.00 | 0.00 | 4.50 |
| 132687056 | 2011.09.12 08:20 | sell | 0.78 | euraud | 1.30807 | 1.39855 | 1.30405 | 2011.09.20 12:43 | 1.33118 | 0.00 | 0.00 | 38.25 | -1 856.83 |
| 132685182 | 2011.09.12 08:11 | sell | 1.53 | euraud | 1.30594 | 1.39640 | 1.30190 | 2011.09.15 17:16 | 1.34627 | 0.00 | 0.00 | 48.66 | -6 357.02 |
| 132641384 | 2011.09.12 00:01 | buy | 1.63 | euraud | 1.30340 | 1.21219 | 1.30669 | 2011.09.12 03:00 | 1.30669 | 0.00 | 0.00 | 0.00 | 558.00 |
| 132641459 | 2011.09.12 00:01 | buy | 0.60 | euraud | 1.30332 | 1.21211 | 1.30661 | 2011.09.12 03:00 | 1.30661 | 0.00 | 0.00 | 0.00 | 205.39 |
| 133746324 | 2011.09.16 12:15 | sell | 0.34 | audusd | 1.03622 | 1.03509 | 1.03189 | 2011.09.16 12:32 | 1.03467 | 0.00 | 0.00 | 0.00 | 52.70 |
| 135452538 | 2011.09.26 18:02 | sell | 1.02 | usdcad | 1.03567 | 1.12585 | 1.03135 | 2011.09.26 18:03 | 1.03548 | 0.00 | 0.00 | 0.00 | 18.72 |
| 135445164 | 2011.09.26 17:40 | sell | 0.99 | usdcad | 1.03564 | 1.12584 | 1.03134 | 2011.09.26 17:41 | 1.03541 | 0.00 | 0.00 | 0.00 | 21.99 |
| 135445432 | 2011.09.26 17:41 | buy | 1.06 | usdcad | 1.03529 | 1.03569 | 1.03958 | 2011.09.26 18:02 | 1.03569 | 0.00 | 0.00 | 0.00 | 40.94 |
| 135443658 | 2011.09.26 17:35 | buy | 0.47 | usdcad | 1.03509 | 1.03570 | 1.03939 | 2011.09.26 17:40 | 1.03570 | 0.00 | 0.00 | 0.00 | 27.68 |
| 133698663 | 2011.09.16 08:42 | buy | 0.26 | audusd | 1.03509 | 1.03542 | 1.03929 | 2011.09.16 12:15 | 1.03618 | 0.00 | 0.00 | 0.00 | 28.34 |
| 135445677 | 2011.09.26 17:42 | buy | 0.81 | usdcad | 1.03481 | 1.03568 | 1.03917 | 2011.09.26 18:02 | 1.03568 | 0.00 | 0.00 | 0.00 | 68.04 |
| 135440622 | 2011.09.26 17:25 | buy | 0.77 | usdcad | 1.03480 | 1.03572 | 1.03914 | 2011.09.26 17:40 | 1.03572 | 0.00 | 0.00 | 0.00 | 68.40 |
| 133698660 | 2011.09.16 08:42 | sell | 0.26 | audusd | 1.03479 | 1.03452 | 1.03059 | 2011.09.16 08:52 | 1.03452 | 0.00 | 0.00 | 0.00 | 7.02 |
| 133749343 | 2011.09.16 12:32 | sell | 0.35 | audusd | 1.03438 | 1.12468 | 1.03018 | 2011.09.19 00:20 | 1.03018 | 0.00 | 0.00 | -6.48 | 147.00 |
| 135441469 | 2011.09.26 17:29 | buy | 0.10 | usdcad | 1.03417 | 1.03429 | 1.03850 | 2011.09.26 17:32 | 1.03429 | 0.00 | 0.00 | 0.00 | 1.16 |
| 128419592 | 2011.08.11 22:58 | sell | 0.10 | audusd | 1.03358 | 1.03333 | 1.02938 | 2011.08.11 23:12 | 1.03333 | 0.00 | 0.00 | 0.00 | 2.50 |
| 133634276 | 2011.09.15 21:09 | buy | 0.23 | audusd | 1.03349 | 0.94319 | 1.03769 | 2011.09.16 08:42 | 1.03484 | 0.00 | 0.00 | 1.82 | 31.05 |
| 133634273 | 2011.09.15 21:09 | sell | 0.23 | audusd | 1.03319 | 1.03310 | 1.02899 | 2011.09.15 22:03 | 1.03310 | 0.00 | 0.00 | 0.00 | 2.07 |
| 133621299 | 2011.09.15 19:41 | buy | 0.22 | audusd | 1.03254 | 1.03288 | 1.03688 | 2011.09.15 20:19 | 1.03288 | 0.00 | 0.00 | 0.00 | 7.48 |
| 133621291 | 2011.09.15 19:41 | sell | 0.22 | audusd | 1.03238 | 1.03206 | 1.02804 | 2011.09.15 20:45 | 1.03206 | 0.00 | 0.00 | 0.00 | 7.04 |
| 133631143 | 2011.09.15 20:45 | buy | 0.23 | audusd | 1.03208 | 1.03318 | 1.03640 | 2011.09.15 21:09 | 1.03318 | 0.00 | 0.00 | 0.00 | 25.30 |
| 133609104 | 2011.09.15 18:25 | sell | 0.16 | audusd | 1.03122 | 1.03118 | 1.02690 | 2011.09.15 18:53 | 1.03118 | 0.00 | 0.00 | 0.00 | 0.64 |
| 133613757 | 2011.09.15 18:53 | buy | 0.21 | audusd | 1.03119 | 1.03198 | 1.03549 | 2011.09.15 19:41 | 1.03235 | 0.00 | 0.00 | 0.00 | 24.36 |
| 133589708 | 2011.09.15 17:13 | buy | 1.89 | audusd | 1.03119 | 1.03123 | 1.03550 | 2011.09.15 18:25 | 1.03123 | 0.00 | 0.00 | 0.00 | 7.56 |
| 128389774 | 2011.08.11 19:58 | sell | 0.41 | audusd | 1.02971 | 1.11992 | 1.02542 | 2011.08.11 19:59 | 1.02907 | 0.00 | 0.00 | 0.00 | 26.24 |
| 134234433 | 2011.09.20 12:53 | buy | 0.26 | audusd | 1.02926 | 1.02994 | 1.03357 | 2011.09.20 13:02 | 1.02994 | 0.00 | 0.00 | 0.00 | 17.68 |
| 128389460 | 2011.08.11 19:56 | buy | 0.57 | audusd | 1.02920 | 1.02989 | 1.03352 | 2011.08.11 19:58 | 1.02989 | 0.00 | 0.00 | 0.00 | 39.33 |
| 128389961 | 2011.08.11 19:59 | buy | 0.39 | audusd | 1.02903 | 0.93881 | 1.03331 | 2011.08.11 22:09 | 1.03331 | 0.00 | 0.00 | 0.00 | 166.92 |
| 128389458 | 2011.08.11 19:56 | sell | 0.57 | audusd | 1.02902 | 1.11920 | 1.02470 | 2011.08.11 19:59 | 1.02874 | 0.00 | 0.00 | 0.00 | 15.96 |
| 128469366 | 2011.08.12 08:39 | sell | 0.15 | audusd | 1.02881 | 1.02813 | 1.02449 | 2011.08.12 08:51 | 1.02813 | 0.00 | 0.00 | 0.00 | 10.20 |
| 128388679 | 2011.08.11 19:51 | sell | 0.47 | audusd | 1.02877 | 1.11892 | 1.02442 | 2011.08.11 19:53 | 1.02836 | 0.00 | 0.00 | 0.00 | 19.27 |
| 128171660 | 2011.08.10 21:12 | buy | 0.20 | audusd | 1.02869 | 1.03000 | 1.03289 | 2011.08.11 19:58 | 1.03000 | 0.00 | 0.00 | 4.56 | 26.20 |
| 128390108 | 2011.08.11 19:59 | buy | 0.42 | audusd | 1.02858 | 0.93837 | 1.03287 | 2011.08.11 22:09 | 1.03287 | 0.00 | 0.00 | 0.00 | 180.18 |
| 128470951 | 2011.08.12 08:54 | sell | 0.15 | audusd | 1.02836 | 1.02818 | 1.02416 | 2011.08.12 09:47 | 1.02818 | 0.00 | 0.00 | 0.00 | 2.70 |
| 128388962 | 2011.08.11 19:53 | buy | 0.52 | audusd | 1.02830 | 1.02853 | 1.03258 | 2011.08.11 19:56 | 1.02900 | 0.00 | 0.00 | 0.00 | 36.40 |
| 128477765 | 2011.08.12 09:48 | sell | 0.14 | audusd | 1.02828 | 1.02524 | 1.02397 | 2011.08.12 10:08 | 1.02524 | 0.00 | 0.00 | 0.00 | 42.56 |
| 128470703 | 2011.08.12 08:51 | sell | 0.17 | audusd | 1.02812 | 1.02781 | 1.02381 | 2011.08.12 10:01 | 1.02781 | 0.00 | 0.00 | 0.00 | 5.27 |
| 128171442 | 2011.08.10 21:11 | sell | 0.23 | audusd | 1.02812 | 1.02795 | 1.02392 | 2011.08.10 21:16 | 1.02739 | 0.00 | 0.00 | 0.00 | 16.79 |
| 128270095 | 2011.08.11 11:35 | sell | 0.18 | audusd | 1.02768 | 1.02739 | 1.02339 | 2011.08.11 12:08 | 1.02677 | 0.00 | 0.00 | 0.00 | 16.38 |
| 128126570 | 2011.08.10 16:49 | buy | 0.13 | audusd | 1.02752 | 1.02814 | 1.03182 | 2011.08.10 21:11 | 1.02814 | 0.00 | 0.00 | 0.00 | 8.06 |
| 128176117 | 2011.08.10 21:42 | buy | 0.13 | audusd | 1.02742 | 1.02807 | 1.03162 | 2011.08.10 21:48 | 1.02807 | 0.00 | 0.00 | 0.00 | 8.45 |
| 128125832 | 2011.08.10 16:45 | buy | 0.13 | audusd | 1.02741 | 1.02741 | 1.03171 | 2011.08.10 16:49 | 1.02741 | 0.00 | 0.00 | 0.00 | 0.00 |
| 128268249 | 2011.08.11 11:27 | buy | 0.14 | audusd | 1.02731 | 1.02742 | 1.03164 | 2011.08.11 11:34 | 1.02742 | 0.00 | 0.00 | 0.00 | 1.54 |
| 128377424 | 2011.08.11 18:51 | buy | 0.26 | audusd | 1.02729 | 1.02829 | 1.03158 | 2011.08.11 19:51 | 1.02877 | 0.00 | 0.00 | 0.00 | 38.48 |
| 128261179 | 2011.08.11 10:41 | buy | 0.29 | audusd | 1.02712 | 1.02790 | 1.03141 | 2011.08.11 11:05 | 1.02790 | 0.00 | 0.00 | 0.00 | 22.62 |
| 128282475 | 2011.08.11 12:35 | buy | 0.13 | audusd | 1.02704 | 1.02791 | 1.03134 | 2011.08.11 17:50 | 1.02791 | 0.00 | 0.00 | 0.00 | 11.31 |
| 128172935 | 2011.08.10 21:16 | buy | 0.39 | audusd | 1.02699 | 1.02744 | 1.03119 | 2011.08.10 21:25 | 1.02744 | 0.00 | 0.00 | 0.00 | 17.55 |
| 128276607 | 2011.08.11 12:08 | buy | 0.14 | audusd | 1.02674 | 1.02691 | 1.03105 | 2011.08.11 12:31 | 1.02691 | 0.00 | 0.00 | 0.00 | 2.38 |
| 128488409 | 2011.08.12 10:21 | sell | 0.18 | audusd | 1.02658 | 1.02635 | 1.02228 | 2011.09.12 21:08 | 1.02635 | 0.00 | 0.00 | -101.11 | 4.14 |
| 132864718 | 2011.09.12 21:28 | sell | 0.61 | audusd | 1.02652 | 1.11682 | 1.02232 | 2011.09.12 22:03 | 1.02618 | 0.00 | 0.00 | 0.00 | 20.74 |
| 132861192 | 2011.09.12 21:08 | buy | 0.55 | audusd | 1.02635 | 1.02652 | 1.03055 | 2011.09.12 21:28 | 1.02652 | 0.00 | 0.00 | 0.00 | 9.35 |
| 132878176 | 2011.09.12 22:16 | sell | 0.77 | audusd | 1.02623 | 1.11653 | 1.02203 | 2011.09.14 06:35 | 1.02203 | 0.00 | 0.00 | -28.57 | 323.40 |
| 128168256 | 2011.08.10 20:52 | buy | 0.47 | audusd | 1.02611 | 1.02664 | 1.03041 | 2011.08.10 21:07 | 1.02710 | 0.00 | 0.00 | 0.00 | 46.53 |
| 132875519 | 2011.09.12 22:03 | buy | 0.70 | audusd | 1.02606 | 1.02629 | 1.03026 | 2011.09.12 22:16 | 1.02629 | 0.00 | 0.00 | 0.00 | 16.10 |
| 128178189 | 2011.08.10 21:58 | sell | 0.10 | audusd | 1.02579 | 1.02576 | 1.02159 | 2011.08.10 22:05 | 1.02576 | 0.00 | 0.00 | 0.00 | 0.30 |
| 128256248 | 2011.08.11 10:25 | buy | 0.21 | audusd | 1.02512 | 1.02560 | 1.02942 | 2011.08.11 10:39 | 1.02560 | 0.00 | 0.00 | 0.00 | 10.08 |
| 128483400 | 2011.08.12 10:08 | sell | 0.10 | audusd | 1.02495 | 1.11514 | 1.02064 | 2011.09.14 08:30 | 1.02064 | 0.00 | 0.00 | -59.89 | 43.10 |
| 134100347 | 2011.09.19 22:24 | sell | 0.37 | audusd | 1.02288 | 1.02271 | 1.01868 | 2011.09.19 23:04 | 1.02271 | 0.00 | 0.00 | 0.00 | 6.29 |
| 134108717 | 2011.09.19 23:04 | sell | 0.37 | audusd | 1.02245 | 1.02242 | 1.01825 | 2011.09.19 23:44 | 1.02242 | 0.00 | 0.00 | 0.00 | 1.11 |
| 134112195 | 2011.09.19 23:45 | sell | 0.37 | audusd | 1.02208 | 1.02149 | 1.01788 | 2011.09.20 00:20 | 1.02149 | 0.00 | 0.00 | -6.70 | 21.83 |
| 134019083 | 2011.09.19 16:03 | buy | 0.41 | audusd | 1.02200 | 1.02204 | 1.02634 | 2011.09.19 22:24 | 1.02273 | 0.00 | 0.00 | 0.00 | 29.93 |
| 134117302 | 2011.09.20 00:20 | sell | 0.37 | audusd | 1.02101 | 1.01937 | 1.01701 | 2011.09.20 01:21 | 1.01937 | 0.00 | 0.00 | 0.00 | 60.68 |
| 134124036 | 2011.09.20 01:21 | sell | 0.38 | audusd | 1.01915 | 1.01798 | 1.01496 | 2011.09.20 01:24 | 1.01769 | 0.00 | 0.00 | 0.00 | 55.48 |
| 134127912 | 2011.09.20 01:27 | sell | 0.92 | audusd | 1.01811 | 1.01742 | 1.01405 | 2011.09.20 01:36 | 1.01726 | 0.00 | 0.00 | 0.00 | 78.20 |
| 134126342 | 2011.09.20 01:24 | sell | 1.97 | audusd | 1.01726 | 1.01695 | 1.01326 | 2011.09.20 01:54 | 1.01695 | 0.00 | 0.00 | 0.00 | 61.07 |
| 134130243 | 2011.09.20 01:45 | sell | 0.67 | audusd | 1.01618 | 1.01556 | 1.01218 | 2011.09.20 02:00 | 1.01545 | 0.00 | 0.00 | 0.00 | 48.91 |
| 134132092 | 2011.09.20 02:01 | sell | 0.89 | audusd | 1.01526 | 1.10576 | 1.01126 | 2011.09.20 12:51 | 1.02879 | 0.00 | 0.00 | 0.00 | -1 204.17 |
| 134132009 | 2011.09.20 02:00 | sell | 1.80 | audusd | 1.01497 | 1.10547 | 1.01097 | 2011.09.20 12:51 | 1.02879 | 0.00 | 0.00 | 0.00 | -2 487.60 |
| 132707059 | 2011.09.12 10:07 | buy | 0.26 | usdcad | 1.00037 | 1.00080 | 1.00467 | 2011.09.12 10:31 | 1.00080 | 0.00 | 0.00 | 0.00 | 11.17 |
| 132707046 | 2011.09.12 10:07 | sell | 0.26 | usdcad | 1.00017 | 0.99783 | 0.99587 | 2011.09.12 12:51 | 0.99783 | 0.00 | 0.00 | 0.00 | 60.97 |
| 132687209 | 2011.09.12 08:21 | buy | 0.39 | usdcad | 0.99999 | 1.00021 | 1.00432 | 2011.09.12 10:07 | 1.00021 | 0.00 | 0.00 | 0.00 | 8.58 |
| 132686428 | 2011.09.12 08:18 | buy | 2.21 | usdcad | 0.99977 | 0.99979 | 1.00407 | 2011.09.12 09:34 | 1.00005 | 0.00 | 0.00 | 0.00 | 61.88 |
| 132767852 | 2011.09.12 14:52 | sell | 0.51 | usdcad | 0.99957 | 0.99854 | 0.99526 | 2011.09.12 15:40 | 0.99843 | 0.00 | 0.00 | 0.00 | 58.23 |
| 132641021 | 2011.09.12 00:00 | buy | 3.01 | usdcad | 0.99767 | 0.90717 | 1.00167 | 2011.09.12 08:11 | 0.99962 | 0.00 | 0.00 | 0.00 | 587.17 |
| 132743251 | 2011.09.12 12:51 | sell | 0.24 | usdcad | 0.99765 | 0.99643 | 0.99333 | 2011.09.12 16:05 | 0.99643 | 0.00 | 0.00 | 0.00 | 29.38 |
| 132641316 | 2011.09.12 00:01 | buy | 2.20 | usdcad | 0.99757 | 0.90698 | 1.00148 | 2011.09.12 08:11 | 0.99967 | 0.00 | 0.00 | 0.00 | 462.15 |
| 132993188 | 2011.09.13 11:22 | buy | 0.40 | usdcad | 0.99710 | 0.90692 | 1.00142 | 2011.09.20 12:40 | 0.98931 | 0.00 | 0.00 | -10.08 | -314.97 |
| 132796818 | 2011.09.12 16:40 | sell | 0.92 | usdcad | 0.99632 | 0.99628 | 0.99202 | 2011.09.12 16:46 | 0.99585 | 0.00 | 0.00 | 0.00 | 43.42 |
| 132786745 | 2011.09.12 16:03 | sell | 0.76 | usdcad | 0.99626 | 1.08641 | 0.99191 | 2011.09.12 16:06 | 0.99598 | 0.00 | 0.00 | 0.00 | 21.37 |
| 132787927 | 2011.09.12 16:06 | sell | 0.93 | usdcad | 0.99584 | 1.08604 | 0.99154 | 2011.09.12 16:51 | 0.99541 | 0.00 | 0.00 | 0.00 | 40.17 |
| 132798515 | 2011.09.12 16:46 | sell | 0.37 | usdcad | 0.99554 | 0.99500 | 0.99121 | 2011.09.12 16:56 | 0.99500 | 0.00 | 0.00 | 0.00 | 20.08 |
| 132986047 | 2011.09.13 11:01 | buy | 0.30 | usdcad | 0.99536 | 0.99695 | 0.99969 | 2011.09.13 11:22 | 0.99695 | 0.00 | 0.00 | 0.00 | 47.85 |
| 132807477 | 2011.09.12 17:14 | sell | 0.45 | usdcad | 0.99517 | 0.99503 | 0.99084 | 2011.09.12 18:48 | 0.99503 | 0.00 | 0.00 | 0.00 | 6.33 |
| 132802101 | 2011.09.12 16:57 | sell | 0.22 | usdcad | 0.99509 | 0.99479 | 0.99079 | 2011.09.12 17:02 | 0.99479 | 0.00 | 0.00 | 0.00 | 6.63 |
| 132805048 | 2011.09.12 17:04 | sell | 0.33 | usdcad | 0.99500 | 0.99480 | 0.99068 | 2011.09.12 17:12 | 0.99480 | 0.00 | 0.00 | 0.00 | 6.63 |
| 132830970 | 2011.09.12 18:48 | sell | 0.10 | usdcad | 0.99494 | 0.99400 | 0.99060 | 2011.09.12 22:53 | 0.99315 | 0.00 | 0.00 | 0.00 | 18.02 |
| 132801968 | 2011.09.12 16:56 | sell | 0.76 | usdcad | 0.99481 | 0.99478 | 0.99051 | 2011.09.12 17:02 | 0.99478 | 0.00 | 0.00 | 0.00 | 2.29 |
| 132804708 | 2011.09.12 17:02 | sell | 0.40 | usdcad | 0.99465 | 1.08482 | 0.99032 | 2011.09.12 22:53 | 0.99414 | 0.00 | 0.00 | 0.00 | 20.52 |
| 132885837 | 2011.09.12 22:53 | buy | 0.43 | usdcad | 0.99412 | 0.99515 | 0.99838 | 2011.09.13 11:01 | 0.99532 | 0.00 | 0.00 | -1.48 | 51.84 |
| 134133532 | 2011.09.20 02:18 | buy | 0.40 | usdcad | 0.99388 | 0.90353 | 0.99803 | 2011.09.20 12:51 | 0.98877 | 0.00 | 0.00 | 0.00 | -206.72 |
| 134130311 | 2011.09.20 01:46 | buy | 0.36 | usdcad | 0.99342 | 0.99353 | 0.99755 | 2011.09.20 02:18 | 0.99353 | 0.00 | 0.00 | 0.00 | 3.99 |
| 132886142 | 2011.09.12 22:53 | buy | 0.33 | usdcad | 0.99310 | 0.99396 | 0.99735 | 2011.09.13 10:16 | 0.99396 | 0.00 | 0.00 | -1.13 | 28.55 |
| 128146471 | 2011.08.10 18:23 | sell | 0.11 | usdcad | 0.99219 | 0.99056 | 0.98790 | 2011.08.10 20:16 | 0.99056 | 0.00 | 0.00 | 0.00 | 18.10 |
| 128143442 | 2011.08.10 18:07 | buy | 0.11 | usdcad | 0.99215 | 0.99225 | 0.99639 | 2011.08.10 18:23 | 0.99225 | 0.00 | 0.00 | 0.00 | 1.11 |
| 128376698 | 2011.08.11 18:47 | buy | 0.21 | usdcad | 0.99084 | 0.90060 | 0.99510 | 2011.09.06 15:48 | 0.99510 | 0.00 | 0.00 | -17.50 | 89.90 |
| 128162725 | 2011.08.10 20:16 | sell | 0.12 | usdcad | 0.99064 | 0.99061 | 0.98650 | 2011.08.10 20:28 | 0.99061 | 0.00 | 0.00 | 0.00 | 0.36 |
| 128131369 | 2011.08.10 17:05 | buy | 0.13 | usdcad | 0.99058 | 0.99089 | 0.99488 | 2011.08.10 17:07 | 0.98902 | 0.00 | 0.00 | 0.00 | -20.51 |
| 128373390 | 2011.08.11 18:30 | buy | 0.24 | usdcad | 0.99053 | 0.99058 | 0.99483 | 2011.08.11 18:44 | 0.99058 | 0.00 | 0.00 | 0.00 | 1.21 |
| 128129908 | 2011.08.10 16:59 | buy | 0.13 | usdcad | 0.99037 | 0.99063 | 0.99467 | 2011.08.10 17:03 | 0.99063 | 0.00 | 0.00 | 0.00 | 3.41 |
| 128371057 | 2011.08.11 18:19 | buy | 0.13 | usdcad | 0.99036 | 0.99048 | 0.99467 | 2011.08.11 18:29 | 0.99048 | 0.00 | 0.00 | 0.00 | 1.57 |
| 128164362 | 2011.08.10 20:28 | sell | 0.12 | usdcad | 0.99029 | 0.99011 | 0.98601 | 2011.08.10 20:38 | 0.99011 | 0.00 | 0.00 | 0.00 | 2.18 |
| 128282610 | 2011.08.11 12:36 | sell | 0.10 | usdcad | 0.99022 | 0.98959 | 0.98598 | 2011.08.11 17:52 | 0.98959 | 0.00 | 0.00 | 0.00 | 6.37 |
| 128131817 | 2011.08.10 17:14 | buy | 0.12 | usdcad | 0.99006 | 0.99101 | 0.99433 | 2011.08.10 18:06 | 0.99149 | 0.00 | 0.00 | 0.00 | 17.31 |
| 128165770 | 2011.08.10 20:38 | sell | 0.12 | usdcad | 0.98992 | 0.98929 | 0.98565 | 2011.08.10 20:47 | 0.98929 | 0.00 | 0.00 | 0.00 | 7.64 |
| 128366905 | 2011.08.11 18:02 | buy | 0.22 | usdcad | 0.98979 | 0.99026 | 0.99417 | 2011.08.11 18:19 | 0.99026 | 0.00 | 0.00 | 0.00 | 10.44 |
| 128281012 | 2011.08.11 12:26 | buy | 0.10 | usdcad | 0.98962 | 0.98968 | 0.99388 | 2011.08.11 12:35 | 0.98968 | 0.00 | 0.00 | 0.00 | 0.61 |
| 128363989 | 2011.08.11 17:53 | buy | 0.23 | usdcad | 0.98945 | 0.99077 | 0.99374 | 2011.08.11 17:58 | 0.99077 | 0.00 | 0.00 | 0.00 | 30.64 |
| 128128158 | 2011.08.10 16:55 | buy | 0.13 | usdcad | 0.98920 | 0.99011 | 0.99348 | 2011.08.10 16:59 | 0.99011 | 0.00 | 0.00 | 0.00 | 11.95 |
| 128167207 | 2011.08.10 20:47 | sell | 0.96 | usdcad | 0.98910 | 0.98909 | 0.98475 | 2011.08.10 20:50 | 0.98877 | 0.00 | 0.00 | 0.00 | 32.04 |
| 134234427 | 2011.09.20 12:53 | buy | 0.34 | usdcad | 0.98909 | 0.89890 | 0.99340 | 2011.09.20 16:36 | 0.99340 | 0.00 | 0.00 | 0.00 | 147.51 |
| 128127363 | 2011.08.10 16:52 | buy | 0.13 | usdcad | 0.98886 | 0.98899 | 0.99312 | 2011.08.10 16:55 | 0.98899 | 0.00 | 0.00 | 0.00 | 1.71 |
| 134233102 | 2011.09.20 12:46 | buy | 0.14 | usdcad | 0.98879 | 0.89866 | 0.99316 | 2011.09.20 12:53 | 0.98889 | 0.00 | 0.00 | 0.00 | 1.42 |
| 128167963 | 2011.08.10 20:50 | sell | 0.42 | usdcad | 0.98863 | 0.98801 | 0.98436 | 2011.08.10 20:51 | 0.98767 | 0.00 | 0.00 | 0.00 | 40.82 |
| 128273564 | 2011.08.11 11:49 | buy | 0.35 | usdcad | 0.98844 | 0.98977 | 0.99276 | 2011.08.11 12:18 | 0.98977 | 0.00 | 0.00 | 0.00 | 47.03 |
| 128126434 | 2011.08.10 16:48 | buy | 0.13 | usdcad | 0.98808 | 0.98864 | 0.99238 | 2011.08.10 16:52 | 0.98864 | 0.00 | 0.00 | 0.00 | 7.36 |
| 128168267 | 2011.08.10 20:52 | sell | 0.34 | usdcad | 0.98797 | 1.07817 | 0.98367 | 2011.08.11 11:40 | 0.98753 | 0.00 | 0.00 | 0.93 | 15.15 |
| 128125804 | 2011.08.10 16:45 | buy | 0.14 | usdcad | 0.98786 | 0.98790 | 0.99209 | 2011.08.10 16:48 | 0.98790 | 0.00 | 0.00 | 0.00 | 0.57 |
| 128271767 | 2011.08.11 11:41 | buy | 0.21 | usdcad | 0.98779 | 0.98781 | 0.99209 | 2011.08.11 11:46 | 0.98781 | 0.00 | 0.00 | 0.00 | 0.43 |
| 128168147 | 2011.08.10 20:51 | sell | 0.46 | usdcad | 0.98763 | 0.98750 | 0.98337 | 2011.08.12 08:41 | 0.98678 | 0.00 | 0.00 | 1.58 | 39.62 |
| 128471242 | 2011.08.12 08:57 | sell | 0.12 | usdcad | 0.98653 | 1.07665 | 0.98215 | 2011.08.15 21:24 | 0.98215 | 0.00 | 0.00 | 0.10 | 53.52 |
| 133594713 | 2011.09.15 17:30 | sell | 1.34 | usdcad | 0.98546 | 1.07582 | 0.98132 | 2011.09.15 20:34 | 0.98519 | 0.00 | 0.00 | 0.00 | 36.72 |
| 129619356 | 2011.08.19 19:31 | sell | 2.01 | usdcad | 0.98517 | 1.07531 | 0.98081 | 2011.08.25 16:31 | 0.98081 | 0.00 | 0.00 | 11.98 | 893.51 |
| 129090512 | 2011.08.17 11:45 | buy | 0.23 | usdcad | 0.98249 | 0.89226 | 0.98676 | 2011.08.18 15:03 | 0.98676 | 0.00 | 0.00 | -2.32 | 99.53 |
| 129086854 | 2011.08.17 11:31 | buy | 0.21 | usdcad | 0.98243 | 0.98256 | 0.98673 | 2011.08.17 11:43 | 0.98256 | 0.00 | 0.00 | 0.00 | 2.78 |
| 128419443 | 2011.08.11 22:57 | sell | 0.97 | nzdusd | 0.83063 | 0.92110 | 0.82660 | 2011.08.12 03:42 | 0.82660 | 0.00 | 0.00 | -8.15 | 390.91 |
| 128419453 | 2011.08.11 22:57 | sell | 0.37 | nzdusd | 0.83062 | 0.92110 | 0.82660 | 2011.08.12 03:42 | 0.82660 | 0.00 | 0.00 | -3.11 | 148.74 |
| 128125769 | 2011.08.10 16:45 | buy | 0.16 | nzdusd | 0.82979 | 0.83014 | 0.83399 | 2011.08.11 22:59 | 0.83014 | 0.00 | 0.00 | 1.49 | 5.60 |
| 128389717 | 2011.08.11 19:58 | buy | 0.52 | nzdusd | 0.82769 | 0.73730 | 0.83180 | 2011.08.11 22:25 | 0.83180 | 0.00 | 0.00 | 0.00 | 213.72 |
| 133741530 | 2011.09.16 11:54 | sell | 0.34 | nzdusd | 0.82662 | 0.91703 | 0.82253 | 2011.09.19 07:24 | 0.82253 | 0.00 | 0.00 | -2.75 | 139.06 |
| 133589784 | 2011.09.15 17:13 | buy | 1.37 | nzdusd | 0.82641 | 0.73600 | 0.83050 | 2011.09.16 11:54 | 0.82663 | 0.00 | 0.00 | 4.66 | 30.14 |
| 133590016 | 2011.09.15 17:14 | buy | 1.08 | nzdusd | 0.82631 | 0.73591 | 0.83041 | 2011.09.16 11:37 | 0.82654 | 0.00 | 0.00 | 3.67 | 24.84 |
| 134234440 | 2011.09.20 12:53 | buy | 0.19 | nzdusd | 0.82491 | 0.82500 | 0.82910 | 2011.09.20 13:01 | 0.82500 | 0.00 | 0.00 | 0.00 | 1.71 |
| 128381162 | 2011.08.11 19:12 | buy | 0.31 | nzdusd | 0.82429 | 0.73382 | 0.82832 | 2011.08.11 19:51 | 0.82571 | 0.00 | 0.00 | 0.00 | 44.02 |
| 132890455 | 2011.09.12 23:12 | buy | 0.58 | nzdusd | 0.82392 | 0.82430 | 0.82794 | 2011.09.13 09:24 | 0.82430 | 0.00 | 0.00 | 1.80 | 22.04 |
| 132961384 | 2011.09.13 09:44 | buy | 0.10 | nzdusd | 0.82385 | 0.82605 | 0.82792 | 2011.09.15 17:12 | 0.82605 | 0.00 | 0.00 | 1.28 | 22.00 |
| 132890498 | 2011.09.12 23:13 | buy | 0.44 | nzdusd | 0.82385 | 0.82423 | 0.82792 | 2011.09.13 09:24 | 0.82423 | 0.00 | 0.00 | 1.36 | 16.72 |
| 128379274 | 2011.08.11 19:00 | buy | 0.29 | nzdusd | 0.82350 | 0.82382 | 0.82766 | 2011.08.11 19:12 | 0.82382 | 0.00 | 0.00 | 0.00 | 9.28 |
| 128376375 | 2011.08.11 18:46 | buy | 0.47 | nzdusd | 0.82310 | 0.82315 | 0.82728 | 2011.08.11 18:56 | 0.82315 | 0.00 | 0.00 | 0.00 | 2.35 |
| 132802026 | 2011.09.12 16:56 | sell | 0.31 | nzdusd | 0.82282 | 0.82244 | 0.81862 | 2011.09.12 18:00 | 0.82244 | 0.00 | 0.00 | 0.00 | 11.78 |
| 128471202 | 2011.08.12 08:57 | sell | 0.14 | nzdusd | 0.82275 | 0.82158 | 0.81860 | 2011.08.12 09:41 | 0.82158 | 0.00 | 0.00 | 0.00 | 16.38 |
| 128469503 | 2011.08.12 08:40 | sell | 0.13 | nzdusd | 0.82273 | 0.82234 | 0.81863 | 2011.08.12 08:50 | 0.82234 | 0.00 | 0.00 | 0.00 | 5.07 |
| 128469357 | 2011.08.12 08:39 | sell | 0.18 | nzdusd | 0.82253 | 0.82231 | 0.81847 | 2011.08.12 08:50 | 0.82231 | 0.00 | 0.00 | 0.00 | 3.96 |
| 128273395 | 2011.08.11 11:48 | buy | 0.45 | nzdusd | 0.82239 | 0.82243 | 0.82656 | 2011.08.11 18:45 | 0.82243 | 0.00 | 0.00 | 0.00 | 1.80 |
| 132886990 | 2011.09.12 22:55 | buy | 0.70 | nzdusd | 0.82237 | 0.82351 | 0.82638 | 2011.09.12 23:12 | 0.82351 | 0.00 | 0.00 | 0.00 | 79.80 |
| 132887013 | 2011.09.12 22:55 | buy | 0.56 | nzdusd | 0.82230 | 0.82350 | 0.82632 | 2011.09.12 23:12 | 0.82350 | 0.00 | 0.00 | 0.00 | 67.20 |
| 132822143 | 2011.09.12 18:08 | sell | 0.77 | nzdusd | 0.82206 | 0.82134 | 0.81790 | 2011.09.12 19:03 | 0.82134 | 0.00 | 0.00 | 0.00 | 55.44 |
| 128470646 | 2011.08.12 08:50 | sell | 0.14 | nzdusd | 0.82200 | 0.82157 | 0.81783 | 2011.08.12 09:41 | 0.82157 | 0.00 | 0.00 | 0.00 | 6.02 |
| 128270514 | 2011.08.11 11:36 | buy | 0.22 | nzdusd | 0.82178 | 0.82190 | 0.82592 | 2011.08.11 11:43 | 0.82190 | 0.00 | 0.00 | 0.00 | 2.64 |
| 134063249 | 2011.09.19 18:52 | sell | 0.38 | nzdusd | 0.82173 | 0.82162 | 0.81757 | 2011.09.19 19:32 | 0.82162 | 0.00 | 0.00 | 0.00 | 4.18 |
| 134019089 | 2011.09.19 16:03 | buy | 0.39 | nzdusd | 0.82170 | 0.82173 | 0.82590 | 2011.09.19 18:52 | 0.82173 | 0.00 | 0.00 | 0.00 | 1.17 |
| 132641730 | 2011.09.12 00:01 | buy | 0.10 | nzdusd | 0.82155 | 0.82159 | 0.82545 | 2011.09.12 16:50 | 0.82159 | 0.00 | 0.00 | 0.00 | 0.40 |
| 128476738 | 2011.08.12 09:41 | sell | 0.16 | nzdusd | 0.82150 | 0.82141 | 0.81732 | 2011.08.12 09:48 | 0.82141 | 0.00 | 0.00 | 0.00 | 1.44 |
| 128477811 | 2011.08.12 09:49 | sell | 0.10 | nzdusd | 0.82142 | 0.81888 | 0.81724 | 2011.08.12 10:08 | 0.81888 | 0.00 | 0.00 | 0.00 | 25.40 |
| 128476727 | 2011.08.12 09:41 | sell | 0.22 | nzdusd | 0.82137 | 0.82131 | 0.81721 | 2011.08.12 09:48 | 0.82131 | 0.00 | 0.00 | 0.00 | 1.32 |
| 128477772 | 2011.08.12 09:48 | sell | 0.11 | nzdusd | 0.82115 | 0.82049 | 0.81698 | 2011.08.12 10:04 | 0.82049 | 0.00 | 0.00 | 0.00 | 7.26 |
| 134069995 | 2011.09.19 19:32 | sell | 0.38 | nzdusd | 0.82114 | 0.82094 | 0.81711 | 2011.09.19 19:42 | 0.82094 | 0.00 | 0.00 | 0.00 | 7.60 |
| 132780645 | 2011.09.12 15:36 | sell | 0.61 | nzdusd | 0.82051 | 0.82019 | 0.81634 | 2011.09.12 19:22 | 0.82019 | 0.00 | 0.00 | 0.00 | 19.52 |
| 128266592 | 2011.08.11 11:05 | buy | 0.22 | nzdusd | 0.82038 | 0.82062 | 0.82450 | 2011.08.11 11:33 | 0.82062 | 0.00 | 0.00 | 0.00 | 5.28 |
| 134071752 | 2011.09.19 19:42 | sell | 0.37 | nzdusd | 0.82033 | 0.81899 | 0.81613 | 2011.09.19 19:58 | 0.81899 | 0.00 | 0.00 | 0.00 | 49.58 |
| 132881538 | 2011.09.12 22:32 | buy | 1.21 | nzdusd | 0.82020 | 0.82188 | 0.82428 | 2011.09.12 22:55 | 0.82188 | 0.00 | 0.00 | 0.00 | 203.28 |
| 132869882 | 2011.09.12 21:43 | buy | 1.33 | nzdusd | 0.82003 | 0.82188 | 0.82411 | 2011.09.12 22:55 | 0.82188 | 0.00 | 0.00 | 0.00 | 246.05 |
| 132837469 | 2011.09.12 19:22 | sell | 0.47 | nzdusd | 0.81999 | 0.81924 | 0.81580 | 2011.09.12 19:43 | 0.81924 | 0.00 | 0.00 | 0.00 | 35.25 |
| 132767935 | 2011.09.12 14:52 | buy | 0.34 | nzdusd | 0.81998 | 0.82001 | 0.82415 | 2011.09.12 15:36 | 0.82049 | 0.00 | 0.00 | 0.00 | 17.34 |

| Ticket | Open Time | Type | Size | Item | Price | S/L | T/P | Close Time | Price | Commission | Taxes | Swap | Profit |
|---|---|---|---|---|---|---|---|---|---|---|---|---|---|
| 134130247 | 2011.09.20 01:45 | sell | 0.49 | nzdusd | 0.81983 | 0.81904 | 0.81577 | 2011.09.20 02:01 | 0.81904 | 0.00 | 0.00 | 0.00 | 38.71 |
| 132837409 | 2011.09.12 19:22 | sell | 0.61 | nzdusd | 0.81979 | 0.81901 | 0.81561 | 2011.09.12 19:42 | 0.81901 | 0.00 | 0.00 | 0.00 | 47.58 |
| 132871244 | 2011.09.12 21:46 | buy | 0.67 | nzdusd | 0.81966 | 0.81977 | 0.82374 | 2011.09.12 22:32 | 0.81977 | 0.00 | 0.00 | 0.00 | 7.37 |
| 128264841 | 2011.08.11 10:58 | buy | 0.10 | nzdusd | 0.81965 | 0.81995 | 0.82378 | 2011.08.11 11:05 | 0.81995 | 0.00 | 0.00 | 0.00 | 3.00 |
| 132842576 | 2011.09.12 19:45 | sell | 0.99 | nzdusd | 0.81953 | 0.81899 | 0.81535 | 2011.09.12 19:52 | 0.81899 | 0.00 | 0.00 | 0.00 | 53.46 |
| 132867715 | 2011.09.12 21:37 | buy | 1.39 | nzdusd | 0.81931 | 0.81970 | 0.82340 | 2011.09.12 21:43 | 0.81970 | 0.00 | 0.00 | 0.00 | 54.21 |
| 132867738 | 2011.09.12 21:37 | buy | 1.08 | nzdusd | 0.81912 | 0.81969 | 0.82320 | 2011.09.12 21:43 | 0.81969 | 0.00 | 0.00 | 0.00 | 61.56 |
| 128263814 | 2011.08.11 10:52 | buy | 0.10 | nzdusd | 0.81906 | 0.81937 | 0.82324 | 2011.08.11 10:57 | 0.81937 | 0.00 | 0.00 | 0.00 | 3.10 |
| 132844389 | 2011.09.12 19:52 | sell | 0.71 | nzdusd | 0.81899 | 0.90932 | 0.81482 | 2011.09.12 20:22 | 0.81869 | 0.00 | 0.00 | 0.00 | 21.30 |
| 128483598 | 2011.08.12 10:09 | sell | 0.10 | nzdusd | 0.81899 | 0.81875 | 0.81482 | 2011.08.19 22:48 | 0.81875 | 0.00 | 0.00 | -5.80 | 2.40 |
| 132842276 | 2011.09.12 19:43 | sell | 1.56 | nzdusd | 0.81896 | 0.90928 | 0.81478 | 2011.09.12 20:22 | 0.81879 | 0.00 | 0.00 | 0.00 | 26.52 |
| 128175855 | 2011.08.10 21:39 | buy | 0.24 | nzdusd | 0.81878 | 0.72829 | 0.82279 | 2011.08.11 05:17 | 0.82279 | 0.00 | 0.00 | 2.23 | 96.24 |
| 132850468 | 2011.09.12 20:22 | buy | 1.16 | nzdusd | 0.81863 | 0.81898 | 0.82282 | 2011.09.12 21:37 | 0.81898 | 0.00 | 0.00 | 0.00 | 40.60 |
| 128483502 | 2011.08.12 10:09 | sell | 0.10 | nzdusd | 0.81859 | 0.81843 | 0.81445 | 2011.08.19 22:48 | 0.81843 | 0.00 | 0.00 | -5.80 | 1.60 |
| 132850833 | 2011.09.12 20:24 | buy | 0.91 | nzdusd | 0.81844 | 0.81896 | 0.82264 | 2011.09.12 21:37 | 0.81896 | 0.00 | 0.00 | 0.00 | 47.32 |
| 128173001 | 2011.08.10 21:17 | buy | 0.10 | nzdusd | 0.81842 | 0.81867 | 0.82250 | 2011.08.10 21:35 | 0.81867 | 0.00 | 0.00 | 0.00 | 2.50 |
| 134074330 | 2011.09.19 19:58 | sell | 0.37 | nzdusd | 0.81832 | 0.81820 | 0.81414 | 2011.09.20 08:53 | 0.81820 | 0.00 | 0.00 | -2.70 | 4.44 |
| 129636620 | 2011.08.19 22:48 | sell | 0.11 | nzdusd | 0.81817 | 0.81677 | 0.81427 | 2011.09.12 08:11 | 0.81628 | 0.00 | 0.00 | -19.84 | 20.79 |
| 128257199 | 2011.08.11 10:28 | buy | 0.10 | nzdusd | 0.81807 | 0.81950 | 0.82219 | 2011.08.11 10:46 | 0.81950 | 0.00 | 0.00 | 0.00 | 14.30 |
| 134132501 | 2011.09.20 02:05 | sell | 0.82 | nzdusd | 0.81794 | 0.81724 | 0.81404 | 2011.09.20 09:13 | 0.81724 | 0.00 | 0.00 | 0.00 | 57.40 |
| 134173333 | 2011.09.20 08:53 | sell | 0.31 | nzdusd | 0.81775 | 0.81724 | 0.81375 | 2011.09.20 09:13 | 0.81724 | 0.00 | 0.00 | 0.00 | 15.81 |
| 128168935 | 2011.08.10 20:59 | buy | 0.10 | nzdusd | 0.81764 | 0.81891 | 0.82176 | 2011.08.10 21:11 | 0.81924 | 0.00 | 0.00 | 0.00 | 16.00 |
| 134177885 | 2011.09.20 09:13 | sell | 0.42 | nzdusd | 0.81709 | 0.90750 | 0.81300 | 2011.09.20 12:52 | 0.82449 | 0.00 | 0.00 | 0.00 | -310.80 |
| 134177867 | 2011.09.20 09:13 | sell | 0.53 | nzdusd | 0.81699 | 0.90740 | 0.81290 | 2011.09.20 12:51 | 0.82449 | 0.00 | 0.00 | 0.00 | -397.50 |
| 132687069 | 2011.09.12 08:20 | buy | 0.51 | nzdusd | 0.81534 | 0.81581 | 0.81937 | 2011.09.12 08:51 | 0.81581 | 0.00 | 0.00 | 0.00 | 23.97 |
| 135465516 | 2011.09.26 18:46 | buy | 0.51 | nzdusd | 0.77229 | 0.68196 | 0.77646 | 2011.09.26 20:55 | 0.77307 | 0.00 | 0.00 | 0.00 | 39.78 |
| 135465886 | 2011.09.26 18:47 | buy | 0.49 | nzdusd | 0.77203 | 0.68171 | 0.77621 | 2011.09.26 20:53 | 0.77234 | 0.00 | 0.00 | 0.00 | 15.19 |
| 135461264 | 2011.09.26 18:28 | buy | 0.43 | nzdusd | 0.77108 | 0.77202 | 0.77525 | 2011.09.26 18:46 | 0.77202 | 0.00 | 0.00 | 0.00 | 40.42 |
| 135442163 | 2011.09.26 17:30 | buy | 0.38 | nzdusd | 0.77095 | 0.77203 | 0.77512 | 2011.09.26 18:46 | 0.77203 | 0.00 | 0.00 | 0.00 | 41.04 |
| 135442201 | 2011.09.26 17:31 | buy | 0.30 | nzdusd | 0.77077 | 0.77078 | 0.77495 | 2011.09.26 18:28 | 0.77078 | 0.00 | 0.00 | 0.00 | 0.30 |
| 135441118 | 2011.09.26 17:27 | buy | 0.39 | nzdusd | 0.77040 | 0.77068 | 0.77459 | 2011.09.26 17:30 | 0.77068 | 0.00 | 0.00 | 0.00 | 10.92 |
| 135440368 | 2011.09.26 17:24 | buy | 1.01 | nzdusd | 0.77031 | 0.77070 | 0.77450 | 2011.09.26 17:30 | 0.77070 | 0.00 | 0.00 | 0.00 | 39.39 |
| 128121926 | 2011.08.10 16:34 | balance | | Deposit | | | | | | | | | 5 000.00 |
| | | | | | | | | | | 0.00 | 0.00 | -208.28 | 7 428.50 |

**Closed P/L:** 7 220.22

**Open Trades:**

| Ticket | Open Time | Type | Size | Item | Price | S/L | T/P | | Price | Commission | Taxes | Swap | Profit |
|---|---|---|---|---|---|---|---|---|---|---|---|---|---|
| 135441451 | 2011.09.26 17:29 | sell | 0.14 | euraud | 1.38461 | 1.38263 | 1.38053 | | 1.38237 | 0.00 | 0.00 | 0.00 | 30.53 |
| 135469663 | 2011.09.26 19:03 | buy | 1.04 | eurusd | 1.34967 | 1.25954 | 1.35404 | | 1.34528 | 0.00 | 0.00 | 0.00 | -456.56 |
| 135482123 | 2011.09.26 20:01 | buy | 0.79 | eurusd | 1.34571 | 1.25559 | 1.35009 | | 1.34528 | 0.00 | 0.00 | 0.00 | -33.97 |
| 135468840 | 2011.09.26 19:00 | buy | 0.77 | gbpusd | 1.55482 | 1.46463 | 1.55913 | | 1.55176 | 0.00 | 0.00 | 0.00 | -235.62 |
| 135471646 | 2011.09.26 19:10 | buy | 1.32 | gbpusd | 1.55374 | 1.46358 | 1.55808 | | 1.55176 | 0.00 | 0.00 | 0.00 | -261.36 |
| 135452541 | 2011.09.26 18:02 | buy | 1.02 | usdcad | 1.03585 | 0.94567 | 1.04017 | | 1.03274 | 0.00 | 0.00 | 0.00 | -307.16 |
| 135452829 | 2011.09.26 18:03 | buy | 0.73 | usdcad | 1.03541 | 0.94529 | 1.03979 | | 1.03274 | 0.00 | 0.00 | 0.00 | -188.73 |
| 135492735 | 2011.09.26 20:53 | sell | 0.58 | nzdusd | 0.77236 | 0.86267 | 0.76817 | | 0.77328 | 0.00 | 0.00 | 0.00 | -53.36 |
| 135493047 | 2011.09.26 20:55 | sell | 0.58 | nzdusd | 0.77318 | 0.86349 | 0.76899 | | 0.77328 | 0.00 | 0.00 | 0.00 | -5.80 |
| | | | | | | | | | | 0.00 | 0.00 | 0.00 | -1 512.03 |

**Floating P/L:** -1 512.03

**Working Orders:**

| Ticket | Open Time | Type | Size | Item | Price | S / L | T / P | Market Price |
|--------|-----------|------|------|------|-------|-------|-------|--------------|

No transactions

**Summary:**

| | | | | | |
|---|---|---|---|---|---|
| **Deposit/Withdrawal:** | 5 000.00 | **Credit Facility:** | 0.00 | | |
| **Closed Trade P/L:** | 7 220.22 | **Floating P/L:** | -1 512.03 | **Margin:** | 8 549.53 |
| **Balance:** | 12 220.22 | **Equity:** | 10 708.19 | **Free Margin:** | 2 158.66 |

**Details:**

| | | | | | |
|---|---|---|---|---|---|
| **Gross Profit:** | 24 497.68 | **Gross Loss:** | 17 277.46 | **Total Net Profit:** | 7 220.22 |
| **Profit Factor:** | 1.42 | **Expected Payoff:** | 14.98 | | |
| **Absolute Drawdown:** | 0.00 | **Maximal Drawdown:** | 14 012.32 (61.51%) | **Relative Drawdown:** | 61.51% (14 012.32) |
| **Total Trades:** | 482 | **Short Positions (won %):** | 263 (93.92%) | **Long Positions (won %):** | 219 (97.72%) |
| | | **Profit Trades (% of total):** | 461 (95.64%) | **Loss trades (% of total):** | 21 (4.36%) |
| **Largest** | | profit trade: | 905.49 | loss trade: | -6 308.36 |
| **Average** | | profit trade: | 53.14 | loss trade: | -822.74 |
| **Maximum** | | consecutive wins ($): | 179 (6 559.55) | consecutive losses ($): | 9 (-6 394.32) |
| **Maximal** | | consecutive profit (count): | 8 312.95 (89) | consecutive loss (count): | -6 394.32 (9) |
| **Average** | | consecutive wins: | 46 | consecutive losses: | 2 |